\documentclass{aa} 

\usepackage{newtxtext,newtxmath} 
\usepackage[switch]{lineno}

\usepackage[T1]{fontenc}
\usepackage{supertabular}
\usepackage{caption}

\DeclareRobustCommand{\VAN}[3]{#2}
\let\VANthebibliography\thebibliography
\def\thebibliography{\DeclareRobustCommand{\VAN}[3]{##3}\VANthebibliography}

\usepackage{graphicx}	
\usepackage{amsmath}	
\usepackage{xcolor}
\definecolor{seagreen}{rgb}{0.180, 0.545, 0.341}
\definecolor{cornflowerblue}{rgb}{0.392, 0.584, 0.929}
\definecolor{coral}{rgb}{1.000 0.498 0.314}
\definecolor{mediumorchid}{rgb}{0.729 0.333 0.827}
\definecolor{darkolivegreen}{rgb}{0.333, 0.420 0.184}
\definecolor{darkkhaki}{rgb}{0.741, 0.718, 0.420}
\definecolor{refereebrown}{rgb}{0.11, 0.11, 0.0}

\begin{document}
\title{Stellar flare detection in XMM-Newton with gradient boosted trees} 

\author{Mario Pasquato
          \inst{1,2}
          \and
          Martino Marelli \inst{1}
                    \and
Andrea De Luca \inst{1} \and
Ruben Salvaterra \inst{1} \and
Gaia Carenini \inst{3,4} \and
Andrea Belfiore \inst{1} \and
Andrea Tiengo \inst{5,1} \and
Paolo Esposito \inst{5} 
          }

\institute{INAF IASF-Milano, Via Alfonso Corti 12, I-20133 Milano, Italy\\
              \email{mario.pasquato@inaf.it}
         \and
             Ciela, Computation and Astrophysical Data Analysis Institute, Montreal, Quebec, Canada\\
             \and
            Département d'Informatique, École Normale Supérieure, Université PSL (Paris Sciences \& Lettres), Paris, France\\ \and
Trinity College, University of Cambridge, Cambridge, UK\\ \and
             IUSS Pavia, Piazza della Vittoria 15, 27100 Pavia, Italy
             }

\date{Received XXX; accepted UUU}


\abstract{The EXTraS project, based on data collected with the XMM-Newton observatory, provided us with a vast amount of light curves for X-ray sources. For each light curve, EXTraS also provided us with a set of features (https://extras.inaf.it). We extract from the EXTraS database a tabular dataset of $31,832$ variable sources by $108$ features. Of these, $13,851$ sources were manually labeled as stellar flares or non-flares based on direct visual inspection.}
{We employ a supervised learning approach to produce a catalog of stellar flares based on our dataset, releasing it to the community. We leverage explainable AI tools and interpretable features to better understand our classifier.}{We train a gradient boosting classifier on $80\%$ of the data for which labels are available. We compute permutation feature importance scores, visualize feature space using UMAP, and analyze some false positive and false negative data points with the help of Shapley additive explanations - an AI explainability technique used to measure the importance of each feature in determining the classifier's prediction for each instance.}{On the test set made up of the remainder $20\%$ of our labeled data, we obtain an accuracy of $97.1\%$, with a precision of $82.4\%$ and a recall of $73.3\%$. Our classifier outperforms a simple criterion based on fitting the light curve with a flare template and significantly surpasses a gradient-boosted classifier trained only on model-independent features. False positives appear related to flaring light curves that are not associated with a stellar counterpart, while false negatives often correspond to multiple flares or otherwise peculiar or noisy curves.}{We apply our trained classifier to currently unlabeled sources, releasing the largest catalog of X-ray stellar flares to date. We estimate that integrating our classifier into the astronomers' workflow will reduce the time spent visually inspecting light curves by approximately half compared to an approach based on flare template fitting, with implications for the classification of sources whose variability is less well established within EXTraS as well as for other catalogs and, possibly, forthcoming missions.}


\maketitle

\section{Introduction} \label{intro}

The EU-FP7 EXTraS project (Exploring the X-ray Transient and variable Sky, \cite{2021A&A...650A.167D}) characterized the aperiodic and periodic variability of all the sources detected by the soft-X-ray (0.2-12 keV) EPIC camera of the ESA telescope {\it XMM-Newton}. The project ended in 2016 and resulted in the full variability characterization of about 500,000 detections obtained before 2012. The systematic EXTraS analysis proved to be very powerful both for discovering peculiar phenomena and for studying samples of sources. For instance \citet{2016A&A...587A..36P} observed a flare from a very early protostar; \citet{2018A&A...616A..36M} detected a very short flare of difficult interpretation from the direction of the Galactic globular cluster NGC6540. 
Moreover, EXTraS proved to be very sensitive, compared to previous investigations of variability: \citet{2021A&A...650A.167D} showed that in a sample of 2357 stars compiled by \citet{2015A&A...581A..28P} a much larger number of variable sources can be found in the EXTraS archive with respect to the 3XMM/4XMM catalogs. 

After the end of the project, we improved the EXTraS pipelines devoted to aperiodic variability. In particular, we generate light curves with uniform time bins by combining data from the three EPIC cameras, which resulted in a higher sensitivity to variability \citep{2022MmSAI..93b.122D}. We have run the updated analysis on all public data up to the end of 2020, obtaining new results such as discovering periodic dips in a peculiar source in M31 \citep[][]{2017ApJ...851L..27M}, determining the orbital period of a fast nova \citep[][]{2018ApJ...866..125M}, investigating the statistical properties of flares from Supergiant Fast X-ray Transients to constrain accretion models for this peculiar class of binary systems \citep[][]{2019MNRAS.487..420S}, and discovering an X-ray superflare from a very cool star of spectral class L1 \citep{2020A&A...634L..13D}.

The systematic analysis of such a large volume of data calls for a machine learning approach, preferably an interpretable one given the need to draw rigorous scientific conclusions from its results. We previously performed a phenomenological classification of all variable sources from the original EXTraS project with an unsupervised approach, which showed that light curves with similar patterns are clustered based on the temporal features computed by the EXTraS analysis \citep{2022A&A...659A..66K}, thus confirming that these features encode useful information for downstream tasks. In this paper, we focus on one such specific classification task, based on a supervised approach: the detection of stellar flares. The study of X-ray flares is a crucial probe of magnetic field generation and dynamics in young stellar objects and, in the main sequence, in cool stars \citep[][]{2015A&A...581A..28P}. It is also very relevant to assess the habitability of extrasolar planets \citep[see e.g.][]{2024LRSP...21....1K}. 

A considerable body of work concerning the application of machine learning to time domain astronomy, including the X-ray band, has accrued over the last decade or so. This includes supervised classification tasks on various precomputed features \citep[][]{2014ApJ...786...20L, 2024ApJ...971..180Y, 2021MNRAS.503.5263Z, 2012ApJ...756...27L, 2024RAA....24h5016Z} as well as unsupervised classification \citep[][]{2022A&A...659A..66K, 2024MNRAS.528.4852P} and, more traditionally, other approaches to automatically detect transients based on statistical methods \citep[e.g.][]{2022A&A...663A.168Q, 2023A&A...675A..44Q, 2019MNRAS.487.4721Y, 2024MNRAS.527.3674R}. In alternative or in addition to statistical learning on human-engineered features, many recent contributions focus increasingly on learning the relevant features directly from the data by means of deep neural networks \citep[][]{2022MNRAS.509.1269O, 2023MNRAS.523.1946R, 2025arXiv250201627S, 2025MNRAS.537..931D}. While we are currently also pursuing such a deep representation learning strategy for light curves and even individual photon detection events (Pasquato et al. in Prep.), the present work focuses on a feature-centric approach.

We trained a gradient boosting classifier \citep[][]{friedman2000additive, friedman2001greedy} to assign flare versus non-flare labels to light curves (LC) based on a subset of their EXTraS features. Training and validation were conducted on $13,851$ manually classified sources. We then applied it to our entire sample of $31,832$ variable light curves in the EXTraS catalog, generated from XMM observations collected between 2000 and 2020, obtaining a predicted class and the associated probability for each source. Crucially, we also pursued humanly intelligible explanations for the behavior of our classifier in the context of the eXplainable AI paradigm \citep[XAI; ][]{gunning2019xai}. Features that are readily understandable by experts in the field let us apply explainability tools more fruitfully. This approach follows the philosophy outlined by \cite{2023arXiv231012528H} and exemplified in previous work by some of us \cite[see e.g.][]{2024ApJ...965...89P}, which establishes explainability as a crucial requirement for machine learning in the natural sciences.

The paper is organized as follows: in Sect.~2 we describe the data and the fiducial sample used to train the model; in Sect.~3 we describe the ML framework; in Sect.~4 we discuss results, to understand the relevance of individual features in the classification and the nature of misclassified instances. We also provide the output catalog of candidate flares. Finally, in Sect.~5 we draw our conclusions.

\section{Data}
\label{sec:data}
In the following, we focus on the features extracted from the intra-observation EXTraS LC (0.2-12 keV flux as a function of time) and derived cumulative distribution functions (CDF; the fraction of time spent by the source below a fixed flux, as a function of the
flux itself) of the point-like sources detected by {\it XMM-Newton/EPIC} between the beginning of 2000 and the end of 2020. These LC have a bin time of 500 s and last from $\sim$1 ks to $\sim$140 ks, depending on the observation time.
Since we want to study their variability, we excluded the least variable LC, defined as those well-fitted (at 5$\sigma$) by a constant model. This resulted in the $31,832$ variable LC data sample used in this paper. It should be noted that the ML model presented in this work, even though it was trained on sources of high variability, is applicable to sources of low variability as well, although likely with a decrease in performance. Low variability sources are not amenable to manual classification though, since the number of sources increases greatly as we lower the cutoff to be considered as variable. 

From each LC and associated CDF, we extracted a number of variability indexes; to choose them, we also took into account the ones used in published machine-learning based papers applied on time series mostly in the X-ray domain: \cite{2011ApJ...733...10R,2014ApJ...786...20L,2015ApJ...813...28F}.
We can divide the variability indexes into three main groups:
\begin{itemize}
    \item model-independent statistical features: these comprehend the most used statistical features (e.g., weighted average, median, skewness, kurtosis) and their associated 1$\sigma$ errors;
    \item model-dependent features: we fitted the LC with a set of models and listed the best-fitted value for each parameter and its associated 1$\sigma$ error and the tail probability; moreover, when possible, the goodness of the fit of different models are compared using an f-test \citep[][]{1969drea.book.....B};
    \item model-independent features of the CDF: by renormalizing both the time and the flux between 0 and 1, we give the coordinates of some specific points of the CDF.
\end{itemize}
From this set of features, we excluded the ones with a straight dependence on the average flux (e.g. the average, the median, the constant feature of every model) since they may bias our model towards brighter sources. This is due to the fact that with the better statistics obtained thanks to higher counts, the type of variability we are looking for is clearer and hence easier to detect. This was already noted during previous work with self-organizing maps \citep[][]{2022A&A...659A..66K}. The potential for bias and the steps taken to mitigate it are discussed in Sect.~\ref{fairness}.

We adopt a final dataset comprising $108$ features for each of our $31,832$ LC, that are the inputs for the supervised analysis. A brief description of the fitting models and of the features is reported in Appendix \ref{label}, while a more complete description of the entire work and its results will be presented in Marelli et al. in prep. 

We trained and tested our machine learning models on a subset of the entire data set (that is, including all features not depending on the average flux) with human-assigned ground truth: our fiducial sample. We generated it by selecting the sources observed between the beginning of 2012 and the end of 2020. We cross-matched on position with optical and multi-wavelength catalogs to select stars, then performed a visual search for flares in LC associated to stars. 

We matched the position of our X-ray sources, as reported in the XMM serendipitous source catalog\footnote{http://xmm-catalog.irap.omp.eu/}, with the position of the stars reported in the SIMBAD catalog\footnote{\url{https://simbad.cds.unistra.fr/simbad}} and in the optical Gaia-DR3 (epoch 2016) catalog \citep{2016A&A...595A...1G,2023A&A...674A...1G}. Using SIMBAD, we considered a star each source cataloged as such, with the exception of the known X-ray binaries (also labeled as HXB, LXB, and/or Pulsar). Using Gaia data, we considered a star each optical source with either a confirmed parallax or a confirmed proper motion at three sigma. We conservatively considered a star each X-ray source that falls within 5" -- the mean positional error of {\it XMM-Newton/EPIC} \citep{2001A&A...365L..18S,2001A&A...365L..27T} -- from a star, as previously defined. The 5" radius we considered for matching is, in fact, generously accounting for both the error attributable to XMM-Newton/EPIC and for the much smaller errors in optical catalogues we matched against. This may lead to concerns that matches to stellar objects may occur serendipitously within the 5" radius even in the absence of a physical association. To check how sensitive our results are to the specific choice of 5" we re-run a cross-match against Gaia with a 2.5" radius. The number of matches (before vetting for parallax) decreases from $24415$ to $21367$, suggesting that most matches occur well within the 5" tolerance, so that the exact value of the radius does not make a major difference. Nonetheless, this conservative approach may lead to certain flares being erroneously tagged as stellar. Since this limitation is impossible to fully overcome even with perfect positional matching, as the example of a star-compact object binary illustrates, we are currently accepting it as a known shortcoming of the training data.

\begin{figure}
\begin{tabular}{c}
\includegraphics[width=\columnwidth]{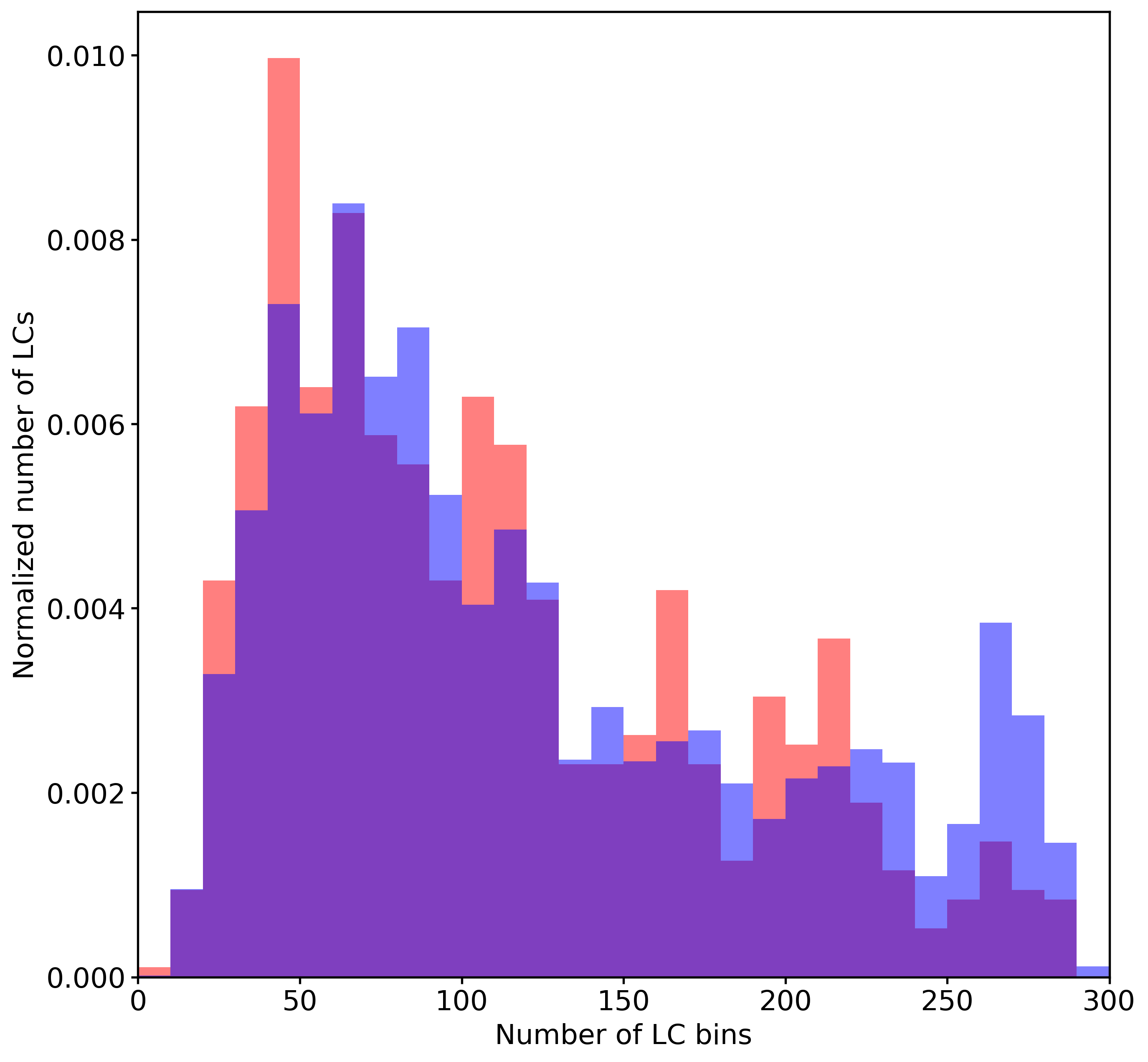}\\
\includegraphics[width=\columnwidth]{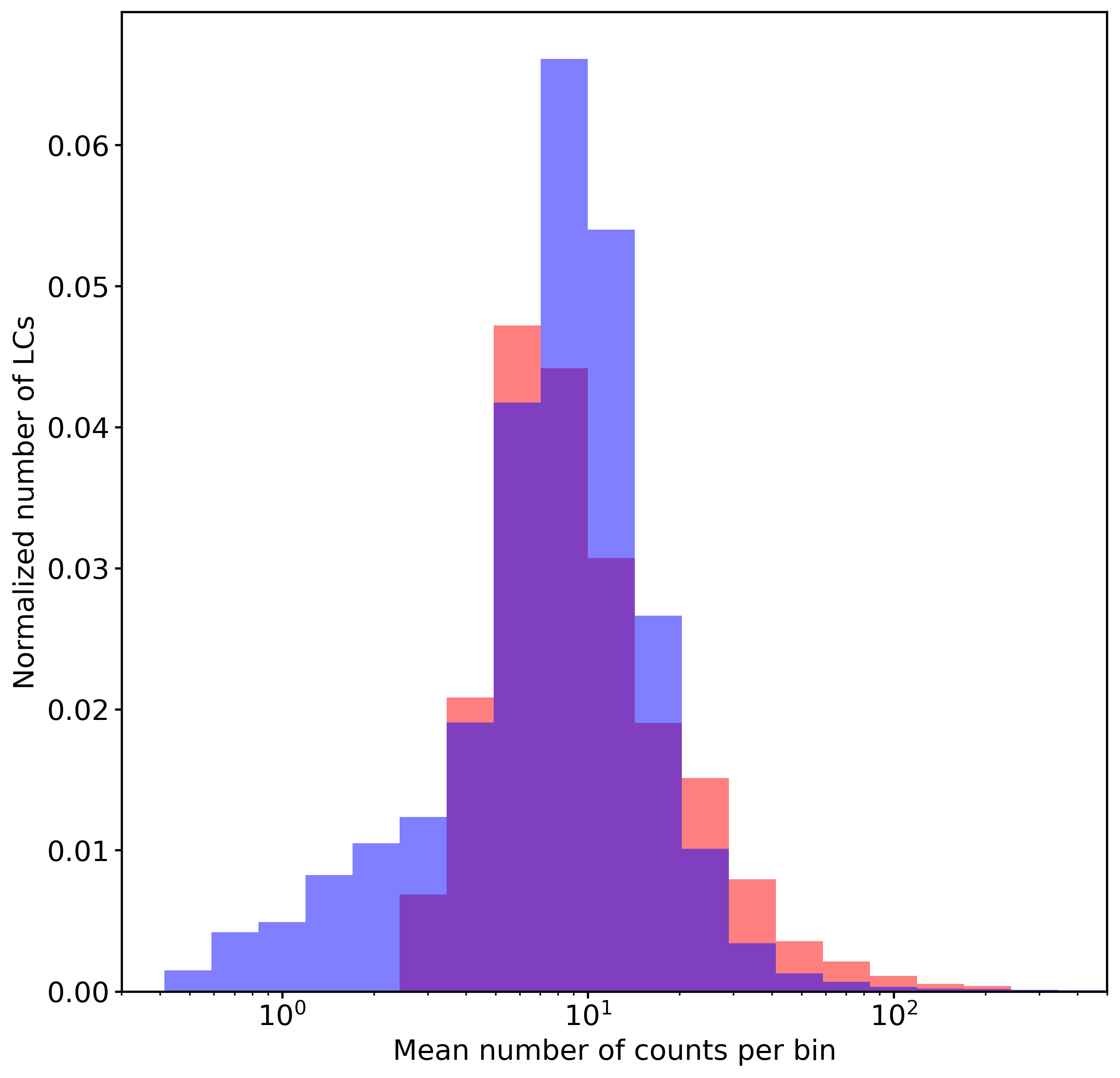}\\
\end{tabular}
\caption{Histograms of the number of bins ({\it Upper Panel}) and mean counts per bin ({\it Lower Panel}) of our sample of $13,851$ light curves. The areas of histograms are normalized to 1. In blue, we show the light curves labeled as "not flaring" and in red the "flaring" ones.\label{figsample}}
\end{figure}

We visually inspected the LC of stars defined as above in search for flares.
Each LC was independently inspected by two different co-authors; each case of disagreement was then discussed, also with the aid of a literature search, if possible (e.g. low-statistics LC containing eclipses only partially covered by the observation could mimic a flare).
We conservatively consider a flare each sudden (within few ks) increase in the LC followed by a sudden or gradual decline. Unfortunately the low statistics of most of the LC make it impossible to differentiate fast-rise exponential-decay (FRED) flares from different types of flares. During the manual labeling phase, the expert coauthors further investigated controversial cases with the help of binning at different time resolutions, as well as adaptively binned LCs following the Bayesian block prescription \citep[][]{2013ApJ...764..167S}. A curve is marked as "flaring" if the variability is dominated by one or more flares. In a fraction of cases, the flare is not entirely comprised within the LC. Such LCs were, however, tagged as flaring when clearly recognized as such.\\
Out of 13,851 LC, we labeled as flaring 953 LC, 163 of which show more than one flare. Our fiducial sample is shown in terms of the observation duration and counts per bin in the histogram in Fig.~\ref{figsample}.

\section{Methods}
\label{sec:methods}
\subsection{Train/test split}
We divided our initial data set into a training and a test set, comprising respectively $80\%$ and $20\%$ of the full dataset. The split was stratified on our label, i.e. the fraction of sources containing a flare was enforced to be roughly the same in the two data sets. We evaluate the performance of our models on unseen test data. 

\subsection{Models}
We trained a gradient-boosted decision-tree model using \texttt{scikit-learn} \citep[][]{scikit-learn} \texttt{GradientBoostingClassifier}. Gradient boosting iteratively builds an additive 
ensemble of weak learners (in our case shallow decision trees) by fitting each new tree to the 
negative gradient (residual) of a specified loss function \citep{friedman2000additive,
friedman2001greedy}.

The procedure starts with a scalar baseline predictor that is constant for every sample. 
For binary classification with logistic loss, this constant equals the log-odds of the 
prevalence of the positive class:
\begin{equation}
  f^{(0)} = \log\!\bigl(\tfrac{1-p}{p}\bigr), 
  \qquad 
  p = \frac{\#\text{positives}}{\#\text{samples}} .
\end{equation}
At this stage no tree is grown; the model outputs the same probability for all observations. 
Each subsequent iteration fits a decision tree $h^{(m)}(\mathbf{x})$ to the current residuals 
and updates the ensemble prediction
\[
  f^{(m)}(\mathbf{x}) = f^{(m-1)}(\mathbf{x}) + \nu\, h^{(m)}(\mathbf{x}),
\]
where $\nu \in (0,1]$ is the learning-rate (shrinkage) factor. Regularization through 
$\nu$, maximum tree depth, and subsampling controls complexity and mitigates over-fitting.

Gradient-boosted trees remain state-of-the-art for tabular data; for instance, 
the XGBoost implementation \citep{2016arXiv160302754C} has repeatedly outperformed more complex 
deep-learning models on diverse benchmarks \citep{2021arXiv210603253S}.

We stuck to the default hyperparameters, avoiding an optimization step that would be computationally intensive and perhaps not particularly valuable at this stage. The only parameter we changed from the defaults was the number of estimators (individual trees grown) which we set to $1000$ rather than the default value of $100$, as a trade-off between improved performance and computing time requirements.

\subsection{Data augmentation}
Our classification task is carried out on an imbalanced dataset, with the positive class representing less than $10\%$ of the instances both in the training and (presumably) in the final deployment data. We have considered mitigating this issue using the Synthetic Minority Oversampling Technique \citep[SMOTE;][]{2011arXiv1106.1813C}. This consists of creating new synthetic instances of the minority class as convex combinations of a given instance and its nearest neighbors. Whether this approach proves successful ultimately depends on the topology of the data. If minority class instances cluster into two or more disconnected but otherwise adjacent regions in feature space, the approach can be detrimental to the performance of a classifier trained on the augmented data. This happens because synthetic instances may end up in a region predominantly populated by instances of the majority class, degrading the precision of the classifier by causing an increase in false positive classifications. For our data set, this issue was empirically confirmed using the SMOTE implementation provided by the library \emph{imbalanced-learn} \citep[][]{2016arXiv160906570L}, which we used with the default settings. We found that a model trained on SMOTE-augmented data where the frequency of the minority class was increased to $20\%$ lost approximately $10\%$ precision compared to the reference model (see Tab.~\ref{metrics}). As a consequence, we avoided using SMOTE to create our final training dataset.

\subsection{Performance metrics}
We evaluate our models in prediction on our test set. The metrics we consider are accuracy, precision, recall, F-score \citep[the harmonic mean of precision and recall; see][]{rijsbergen1979information}, as well as the whole precision-recall curve. 
Accuracy is the fraction of correctly classified samples over the total. Raw accuracy is not a particularly useful metric in this context, given how unbalanced the classes are in our data set: as we discuss below, an accuracy exceeding $90\%$ can be obtained merely be predicting the majority class (no flare). We still report our accuracy for each model for the sake of completeness.

Precision and recall are computed for the flare class. The former corresponds to the fraction of actual flares (true positives) over the total number of sources classified as flares. This metric is also called `purity' in the context of astronomical catalogues. Recall is the fraction of true positives over the total number of sources that are actually flares. This metric is also called `completeness' in the context of astronomical catalogues. Unlike accuracy, precision is crucially important for the goals of our classification, which is to produce a sample of candidate sources that are enriched in actual flares, to reduce the time devoted to human visual inspection for a given number of confirmed flares. Recall is also important, but secondary to precision, since the former immediately translates to savings in human classification work.

Typically, and in our case in particular, a classifier returns the predicted probability for each data-point to be a member of a given class. The final binary classification requires setting a threshold on such probability so that a given data-point is predicted as a member if its predicted probability of membership exceeds the threshold. Only then can a value of precision, recall, and accuracy be obtained. The choice of the threshold results in a tradeoff between the risk of false positive results and that of false negative results. As such this choice depends on the relative costs of false positives and false negatives. It is thus common practice to evaluate a classifier at multiple thresholds, obtaining a precision-recall curve which is an overall quantification of the performance of the classifier. 

\subsection{Explainability}
Gradient boosting models, being based on an ensemble of decision trees, are not inherently interpretable. 
Using this kind of models is justified given our goal of sifting a large amount of data to reduce manual classification labour, since the ultimate decision on follow-up observations and similar expensive actions will be still guided by human visual inspection of the relevant light curves. However, seeking explanations for the behavior of our models is helpful in, on the one hand, building confidence in their predictions and on the other, iteratively improving the models. 

We thus leverage XAI techniques to gain insight into our model's behavior. In particular, we use permutation feature importance \citep[PFI; ][]{altmann2010permutation}, and Shapley additive explanations \citep[SHAP][]{lundberg2017unified}. The former measures the impact of each feature on the model's prediction accuracy by randomly permuting the feature -thus making it useless for prediction- and observing the resulting decrease in accuracy, while the latter is a game theoretic strategy introduced originally by \cite{shapley1953value} to equitably divide earnings in cooperative games, that has been applied to assign a value to each feature based on its contribution to the prediction on a given data point when it is combined with other features.
This method considers all possible combinations\footnote{Computational considerations permitting.} of features as player coalitions and computes the marginal contribution of each feature to the prediction outcome by comparing the prediction with and without the given feature. The Shapley value for a feature is then calculated as the weighted average of its marginal contributions across all possible coalitions. This approach ensures a fair attribution of the prediction outcome to individual features, adhering to properties such as efficiency (the total contribution of all features equals the total change in prediction from a baseline prediction), symmetry (features with identical contributions receive identical Shapley values), dummy (features that do not change the prediction receive a Shapley value of zero), and additivity (the Shapley values for a model that is a sum of several models equal the sum of the Shapley values of each model). Shapley values, unlike PFI, can deal with interactions between features because it considers all possible coalitions.

We also visualize the space of the most important features (according to PFI) using Uniform Manifold Approximation and Projection \citep[UMAP;][]{2018arXiv180203426M}. UMAP leverages a graph representation of the data in the high-dimensional input space, by connecting each point to its nearest neighbors. The number of points considered in this step is adjustable by the user and influences the final outcome by privileging global over local structure as it is increased. Having built a suitable graph, UMAP then  seeks to embed this graph in a lower-dimensional space in a way that best preserves faithfulness to the original data. This process involves minimizing the cross-entropy between two similar fuzzy sets representing the high-dimensional and low-dimensional spaces, respectively.

We run PFI $50$ times, thus obtaining mean importances and their standard deviations over the $50$ runs. This allows us to select the features that are important at one standard deviation, which yields six features. UMAP was run on the space of these six features after each one has been scaled using the robust scaler preprocessing tool (from \emph{scikit-learn} preprocessing) so that the first quartile of the original feature corresponds to $0$ in the scaled feature and the third quartile to $1$. A tedious manual exploration of the results of adopting different hyperparameters for UMAP was carried out while searching for insight into our data set. In particular we tweaked the \emph{n\_neighbors} and \emph{min\_dist} hyperparameters in umap-learn's implementation of UMAP to avoid the formation of "horseshoes", winding, elongated one-dimensional structures that appear when a dimension reduction method recognizes an essentially one-dimensional structure in a section of the data \citep[][]{2008arXiv0811.1477D}. These may arise even in the absence of a genuinely one-dimensional manifold in the data-generating process, e.g. due to overly local connectivity. Further, we varied umap's hyperparameters to confirm that certain features on the map, in particular islands and peninsulas to which we attached a meaning as discussed in Sect.~\ref{appe}, were stable. 
As an additional way to check that the features selected by PFI are indeed capturing most of the useful signal for our classification task, we retrained a model on these features only, obtaining a limited reduction in performance.

Finally, we use individual conditional expectation (ICE) curves to understand how the model predictions depend on each feature. ICE plots are a visualization technique used to analyze the effect of a single feature on the predictions made by a machine learning model, for a sample of observations from the dataset. An ICE plot does this by varying the value of the feature of interest across its range and observing how the prediction changes, while keeping all other features fixed. This process is repeated for every LC, resulting in a series of lines on the plot—one for each LC. These lines show how the predictions would change if only that one feature were altered, providing insight into the feature's individual effect on the model output. ICE plots are particularly useful for identifying how the relationship between a feature and the prediction varies across different instances in the dataset, showcasing the heterogeneity in the model dependence on that feature.

\subsection{Bias mitigation}
\label{fairness}
Machine learning models will use any information available in the data to predict the labels, even when a human would understand that certain information is better left out. Several features correlate with overall source brightness, but we do not want our model to more confidently classify a bright source as a flare just because the data shows that a flare pattern is more easily spotted by human experts in a bright source than in a dim one. The appropriate theoretical framework for discussing this kind of biasing effect introduced by certain features is that of fairness in machine learning \citep[see][for a review]{10.1145/3616865}.

Fairness in classification is commonly formalized with respect to a
\emph{protected attribute} \(A\)  that the classifier should not discriminate on.  
A simple statistical notion of fairness is demographic parity, requiring that the distribution of predictions be the same between groups, independently of the true label:

\begin{equation}
    \forall s,\; \forall a,b:\quad
    P(S = s \mid A = a) \;=\; P(S = s \mid A = b).
\end{equation}

For a given definition of fairness, departure from these objectives counts as evidence of bias.

There are several mitigation strategies, the most intuitive of which is so-called \emph{fairness by unawareness}, consisting in removing the protected attribute from the training features. In our case, features such as AVE where not included in the training set. However this approach in not sufficiently effective: a machine learning model will latch onto any other feature that correlates with the protected attribute if that leads to improved predictions.

We have thus defined our protected attribute as the new binary feature BRIGHT, obtained by thresholding AVE on its median. The hand-labeled data show a mild but significant difference in the prevalence of flares among BRIGHT$=$True and BRIGHT$=$False sources: $8.3 \%$ against $7.0\%$ respectively. We used the fairlearn library \citep[][]{bird2020fairlearn} to test a correlation-removal pre-processing mitigation strategy for our dataset in addition to fairness by unawareness.

\section{Results}
\label{sec:results}
\subsection{Performance on test data}

The accuracy, precision, recall, and F-score for our models on our test set (which was withheld in training) are presented in Tab.~\ref{metrics}. We compare our full model (first row in the table), trained on all available features, to a model trained on a subset of features that excludes features obtained from model-dependent and computationally intensive fits of physically motivated templates to LC (second row) and to a model trained only on the most important features according to PFI (listed in Tab.~\ref{pfi}; last row). To compare the performance of our model with a simple human-based method, we do not have a standard baseline. In order to search for flares we usually make a cut on F\_NSIGMA\_FLCON\footnote{A multi-component cut using other flare model parameters does not result in a significant improvement, or does introduce clear biases}, that is the probability coming from an f-test \citep{1969drea.book.....B} performed comparing a constant+FRED model with a constant model\footnote{We report the probability in terms of number of sigmas in order to normalize the distribution of the values, thus a low probability turns in a high value of F\_NSIGMA\_FLCON that show that the improvement by adding the flare model to the constant is hardly by chance.}. Thus, in order to search for real flares, we cut out every LC with a F\_NSIGMA\_FLCON below a certain threshold, usually 5. This is expected to be the best parameter to cut, since, by definition, it indicates the improvement by adding a flare model even in the case of a multi-component LC, while the parameter associated with the goodness of the constant + FRED flare (FL\_NSIGMA) would fail. Thus, we compare our models with this cut (third row in Tab.~\ref{metrics}). Notice how the full model outperforms every other in terms of accuracy and precision, while the baseline flare fit yields a high recall at the expense of precision, which is an abysmal $33.4\%$. Moreover, it is possible to change the classification threshold of our full model to obtain the same level of recall as the baseline flare fit, while keeping a precision of $66.0\%$, which is still roughly double the baseline.

\begin{figure}
\includegraphics[width=\columnwidth]{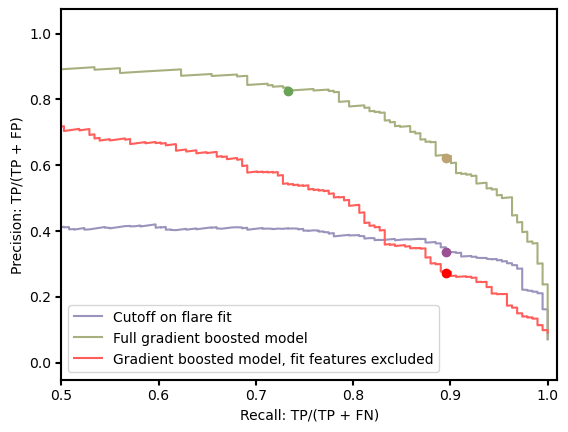}
\caption{Precision-recall curves for setting a cutoff on F\_NSIGMA\_FLCON (purple), with the brighter purple dot corresponding to a five-sigma cutoff. The green curve corresponds to our best gradient boosting model (including all features), with the brighter green dot corresponding to a threshold of $0.5$. The brown dot on this curve corresponds to the level of precision that would be reached at the higher recall reached by the five-sigma cutoff on F\_NSIGMA\_FLCON. The red curve corresponds to a gradient boosting model including only features that do not derive from fitting physically meaningful templates to the light curve (such as flares, dips, eclipses, etc.), and the brighter red dot corresponds to the level of precision that would be reached at the recall reached by the five-sigma cutoff on F\_NSIGMA\_FLCON.\label{prc}}
\end{figure}

The tradeoff between precision and recall is best appreciated in Fig.~\ref{prc} where we present three precision-recall curves obtained by varying the relevant thresholds for each classification method, while the results reported in Tab.~\ref{metrics} correspond to a conventional threshold of $0.5$. The green curve represents our gradient boosted classifier including all $108$ features, the red curve the classifier trained only on model-independent features, and the purple curve the cut on the F\_NSIGMA\_FLCON parameter (in this case, the thresholds are the different possible values at which we apply the cut).

Each point on any of these lines represents a value of precision and recall obtained by changing said threshold, without altering the underlying model. A curve situated towards the upper right of the plot represents a higher overall performance.

The classifier trained on all $108$ features clearly outperforms the others at all recall levels. Excluding model-dependent features strongly reduces the performance, although at sufficiently low recall (below $\approx 0.835$) it still performs better than a naive cutoff on F\_NSIGMA\_FLCON. 

\begin{table}
\caption{Metrics for our models. Full model trained on all features (first row), restricted model trained on a subset of model-independent features (second row), baseline relying on a flare template fit (third row), baseline predicting always the majority class (fourth row), model restricted to only the most PFI-important features (fifth row), and model trained SMOTE-augmented dataset (sixth row).\label{metrics}}
\begin{tabular}{lllll}
\hline
\hline
Model & Accuracy & Precision & Recall & F-score\\
\hline
Full & $97.1\%$ & $82.4\%$ & $73.3\%$ & $77.6\%$\\
No fit & $95.0\%$ & $69.4\%$ & $48.7\%$ & $57.2\%$\\
Cutoff & $87.0\%$ & $33.4\%$ & $89.5\%$ & $48.6\%$\\
Majority & $93.1\%$ & NA & $0.0\%$ & NA\\
Important feat. & $96.2\%$ & $77.8\%$ & $70.9\%$ & $74.2\%$\\
SMOTE & $95.6\%$ & $69.2\%$ & $77.9\%$ & $73.3\%$ \\
\hline
\end{tabular}
\end{table}

\subsection{The role of important features}
\label{appe}
Tab.~\ref{pfi} presents the PFI importance for the most important features. These have been selected by repeating the permutation procedure $50$ times and obtaining mean and standard deviation values of the PFI for each feature, including in the final table only those for which the mean differs from zero by at least one standard deviation. 

\begin{table}
\caption{Permutation feature importance for the features that are important at least at one sigma. Features are marked as model dependent if they are the result of fitting a physically motivated template to the LC.\label{pfi}}
\begin{tabular}{lll}
\hline
\hline
Feature & Importance & Model-dependent?\\
\hline
F\_NSIGMA\_FLCON & $0.059  \pm 0.004$ & Yes\\
FL\_NSIGMA & $0.010 \pm 0.002$ & Yes\\
FL\_DT\_ERR & $0.009 \pm 0.002$ & Yes\\
MEDMAXOFF & $0.005 \pm 0.002$ & No\\
FLUX\_P50 & $0.0006 \pm 0.0004$ & No\\
TFRAC\_BEL1S & $0.0005 \pm 0.0003$ & No\\
\hline
\end{tabular}
\end{table}

The features deemed important include F\_NSIGMA\_FLCON and FL\_NSIGMA, two parameters that are related to the goodness of fit with a flare template, as well as FL\_DT\_ERR, quantifying the error on the decline time of the flare fit. Additionally, MEDMAXOFF, FLUX\_P50, and  TFRAC\_BEL1S further characterize the variability of the source in a model independent way. It is unsurprising that F\_NSIGMA\_FLCON and FL\_NSIGMA appear as important features, given the fact that these statistics are designed precisely to select the curves that most resemble a flare template. Similarly, FL\_DT\_ERR represents the error in the decay time of the flare template fit; it is clear that a large error may indicate a poor fit, and may also appear in cases where the rise in flux happens close to the end of the observation. Those are most likely spurious. The role of MEDMAXOFF is also immediately evident: a large difference between the maximum and the median of a light curve signifies a sudden spike that decays rapidly.

To gain further insight into the role of this subset of most important features with respect to the others, we trained a new model only on them. As shown in Tab.~\ref{metrics}, the full model outperforms the restricted model, but not by a huge margin: raw accuracy drops by about one percentage point. While the six most important features are sufficient to constrain predictions on the bulk of the data, it appears likely that edge cases have to rely on less frequently used features. In fact this is confirmed by our subsequent analysis of instance-level predictions, on false-positives and false-negatives.

\begin{figure}
\includegraphics[width=\columnwidth]{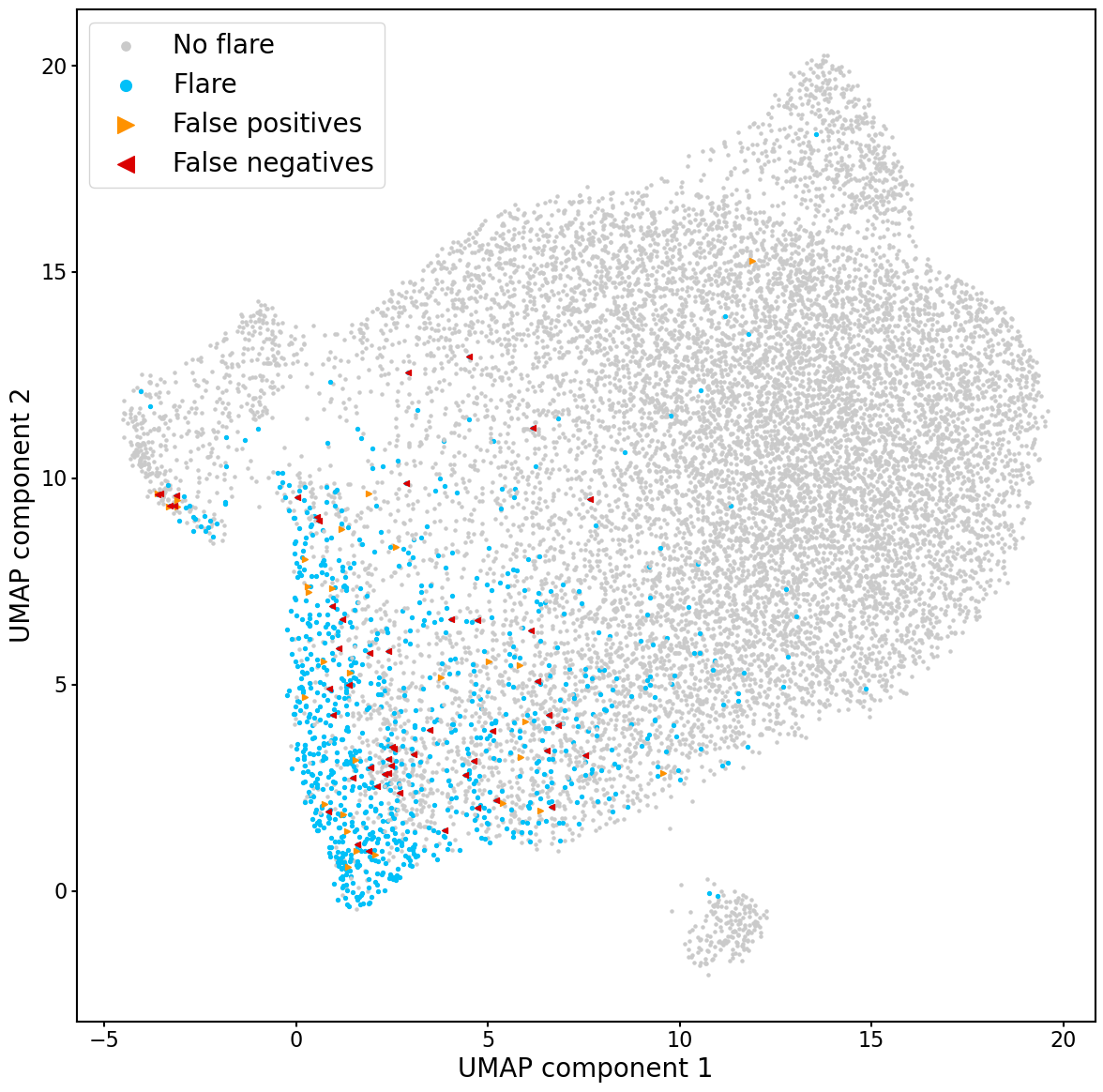}
\caption{UMAP embedding calculated on the important features from the training set (flares in cerulean blue, non-flares in light gray), test set displayed in the same coordinates. False positives within the test set are shown in orange and false negatives in red.\label{umapplot}}
\end{figure}

\begin{figure}
\includegraphics[width=\columnwidth]{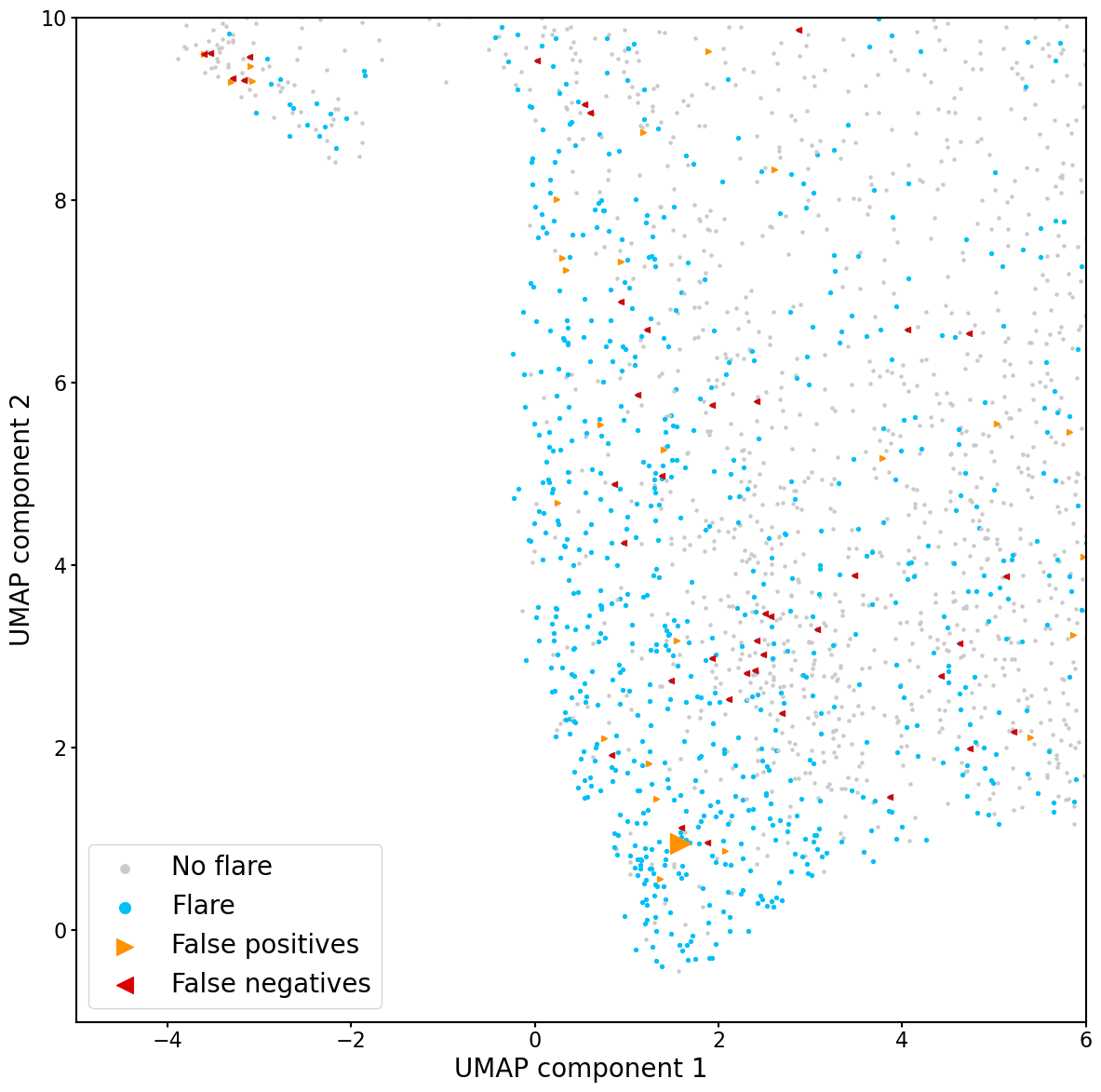}
\caption{UMAP embedding calculated on the important features from the training set (flares in cerulean blue, non-flares in light gray), test set displayed in the same coordinates. False positives within the test set are shown in orange and false negatives in red. Zoom on flareland, showing the false positive source $0722360301/5$ (discussed below) with a bigger triangle.\label{umapplotzoom}}
\end{figure}

To better understand the relationships between features and their interactions in prediction, we explored the entire six-dimensional space of the features that are important at one sigma to identify where flares are preferentially located with respect to relevant substructures such as clustering of similar sources. This was carried out by means of visualization with UMAP, which allowed us to represent our six-dimensional data points on the plane, as shown in Fig.~\ref{umapplot}, where we also mark actual flares, false positives, and false negatives. Individual features are visualized on the same plane in Appendix~\ref{appendicebonus}. Two of the most prominent features, the fin-like peninsula at the top right and the small island at the bottom, are related to extreme values of FL\_DT\_ERR: unrealistically large (much longer than the typical duration of an observation) in the former, and exactly zero in the latter. Both values indicate that a flare fit essentially failed, and in fact the top peninsula contains no flares and the bottom island only two.
The peninsula at the top left is associated to high values of FL\_N\_SIGMA, even though it does not contain many flares: these are sources that are mostly not stellar flares, yet have been well fit by a flare template. These may be curves with a step increase in flux, which end up well fitted by flare models with a long decay time.
At the bottom-left we find flareland: there F\_N\_SIGMA\_FLCON and MEDMAXOFF are high, while FL\_DT\_ERR and FLUX\_P50 are low. Flareland is shown at a closer level of zoom in Fig.~\ref{umapplotzoom}.

\begin{figure}
\includegraphics[width=\columnwidth]{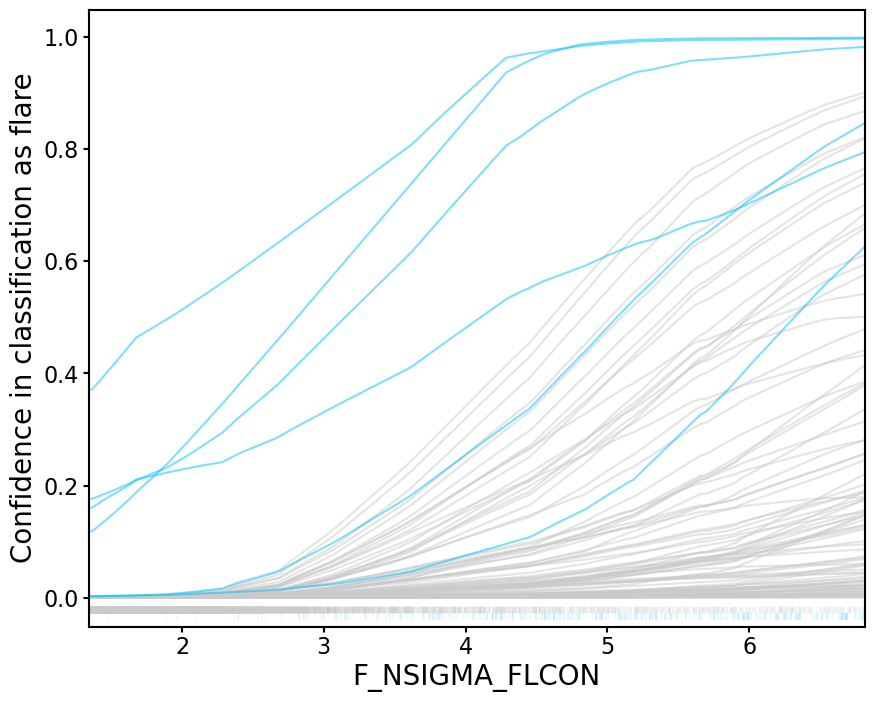}
\caption{ICE plot for F\_NSIGMA\_FLCON. Curves for $100$ randomly chosen sources are shown. Non-flare sources are shown in gray, flares in cerulean blue. At the bottom a rug plot shows the actual values of the feature taken on by flares (cerulean blue) and non-flares (gray). \label{ice0}}
\end{figure}

A complementary way to understand how our machine learning model is using our features is ICE curves.
Fig.~\ref{ice0} shows that the full model is strongly relying on the flare fit result to find flares, as expected. The predicted probability of being predicted as a flare increases following a sigmoid-like curve, that is slowly at first, then roughly linearly with FL\_NSIGMA\_FLCON up until almost $1$, where it flattens out. The ICE curves are mostly non-intersecting and similar in shape, suggesting that FL\_NSIGMA\_FLCON does not have major interactions with other features.

\begin{figure}
A\includegraphics[width=\columnwidth]{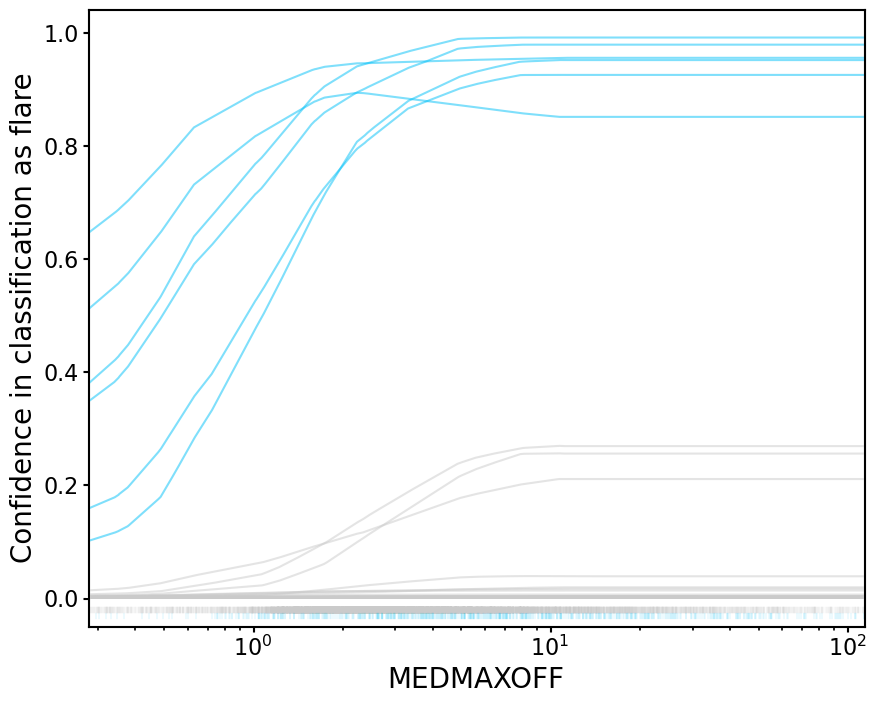}
\caption{ICE plot for MEDMAXOFF. Curves for $100$ randomly chosen sources are shown. Non-flare sources are shown in gray, flares in cerulean blue. At the bottom a rug plot shows the actual values of the feature taken on by flares (cerulean blue) and non-flares (gray). \label{ice3}}
\end{figure}

As a second example of feature whose meaning gets clarified by an ICE plot, we show MEDMAXOFF in Fig.~\ref{ice3}. A flare is a sudden, isolated increase in flux, so a large offset between the median and the maximum of the light curve is a signal that a source may be flaring. This is however per se not a sure indication of a flare, hence for most data points the response shown by the ICE curve is flat (bottom part of the image), while the model relies on MEDMAXOFF only for points that, due to other features, are already suspect flares. This can be understood as an example of interaction between features.

\subsection{Understanding misclassified instances} \label{sec:miscl}
\begin{figure}
\includegraphics[width=\columnwidth]{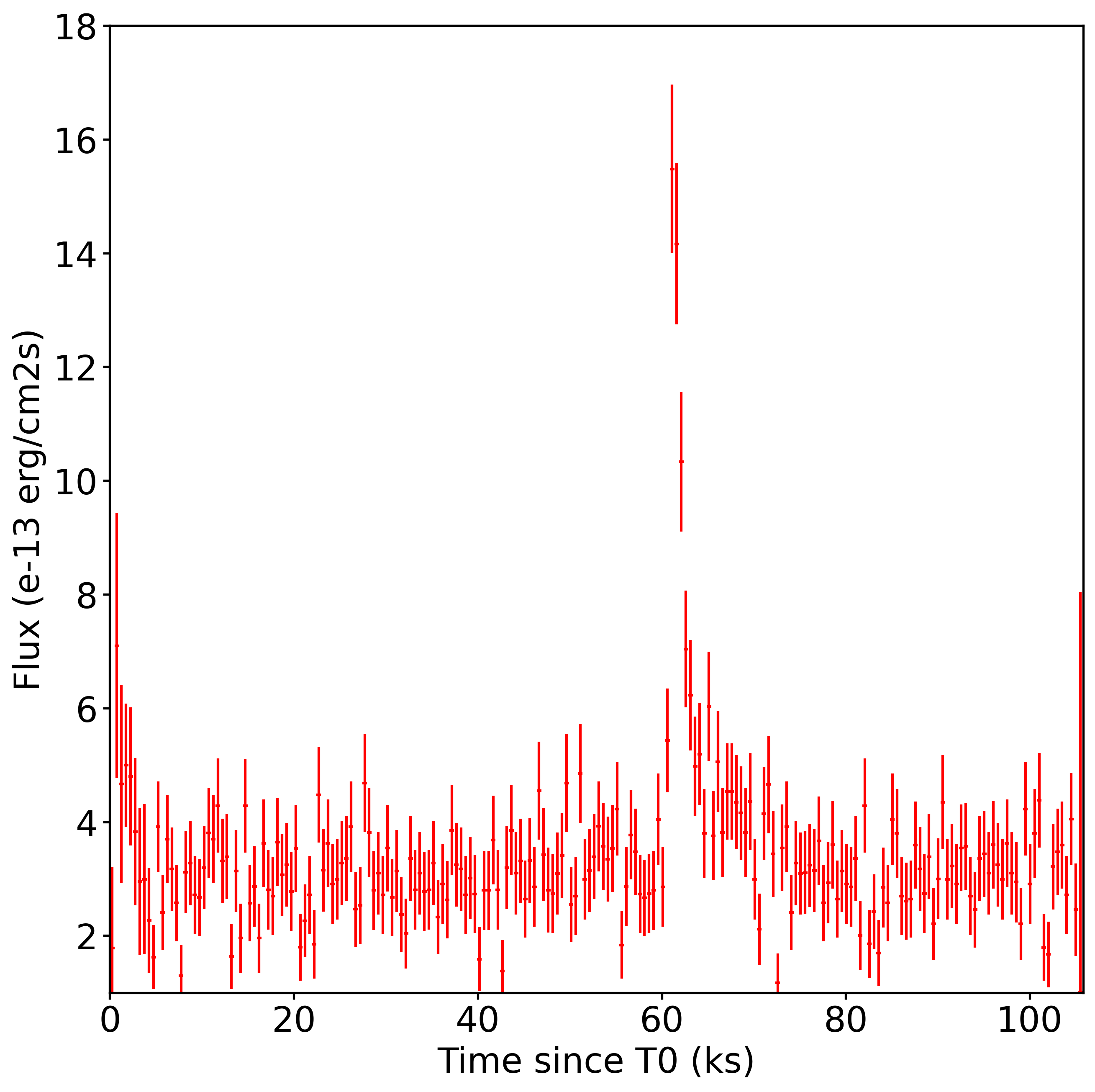}\\
\caption{Light curve for $0722360301/5$. An apparent flare, but it is not associated with a star. T0 is the time of the first photon of the observation. \label{fapocurve}}
\end{figure}

\begin{figure}
\includegraphics[width=\columnwidth]{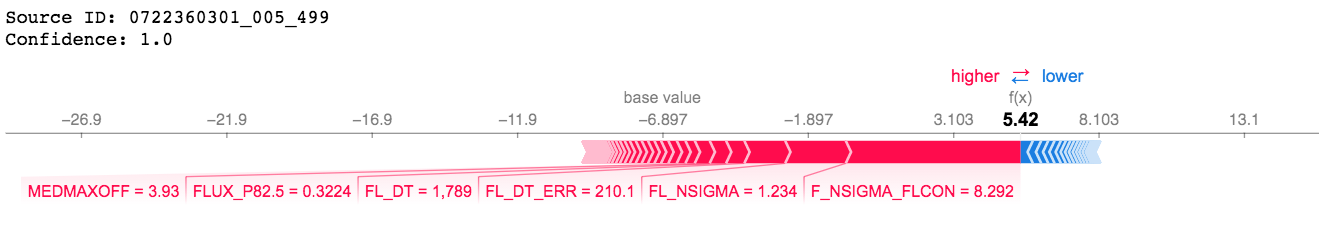}
\caption{Shapley values for a paradigmatic false positive source, $0722360301/5$.\label{fapo}}
\end{figure}

Our classifier misclassifies $81$ LC out of $2771$. It is natural to wonder why such misclassification occurs. In Fig.~\ref{fapocurve} we show a paradigmatic misclassified source among the false positives -for which the model predicts the flare class but the ground truth label has no flare. Ground truth labels were assigned by human inspection to flare patterns that corresponded to a stellar counterpart. However flare-like activity can be associated also to non-stellar sources or to stars that are undetected in current surveys, and this is likely often the case for false positives: Fig.~\ref{fapocurve} in particular appears to be a clear cut flare -in fact Fig.~\ref{umapplotzoom} shows it deeply embedded in flareland- but it did not have a stellar counterpart. Consequently, it has not been labeled as a stellar flare. Interestingly, SHAP shows (Fig.~\ref{fapo}) the source was in fact classified as such because of variables related to the flare template fit, whose role in classification we have clarified above. Shapley values thus suggest that false positives may be flares not associated to a stellar source. This is a limitation of the training set rather than of the ML model: a purely phenomenological definition of flare would likely have resulted in a lower incidence of false positives, but we chose to label only confirmed stellar flares as positive instances in our training data to maximize the usefulness of the catalog for scientific purposes.

\begin{figure}
\begin{tabular}{c}
\includegraphics[width=0.8\columnwidth]{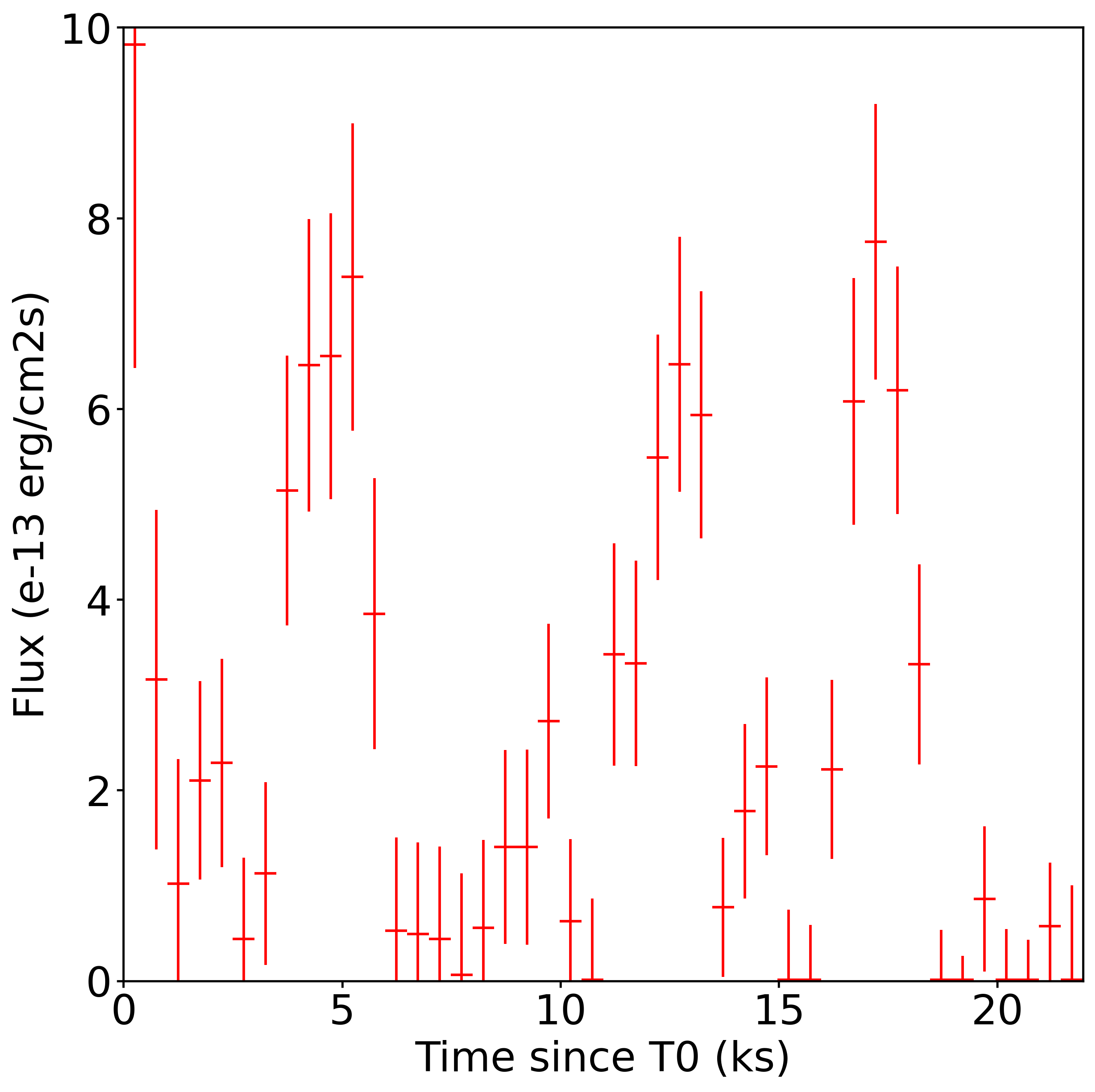}\\
\includegraphics[width=0.8\columnwidth]{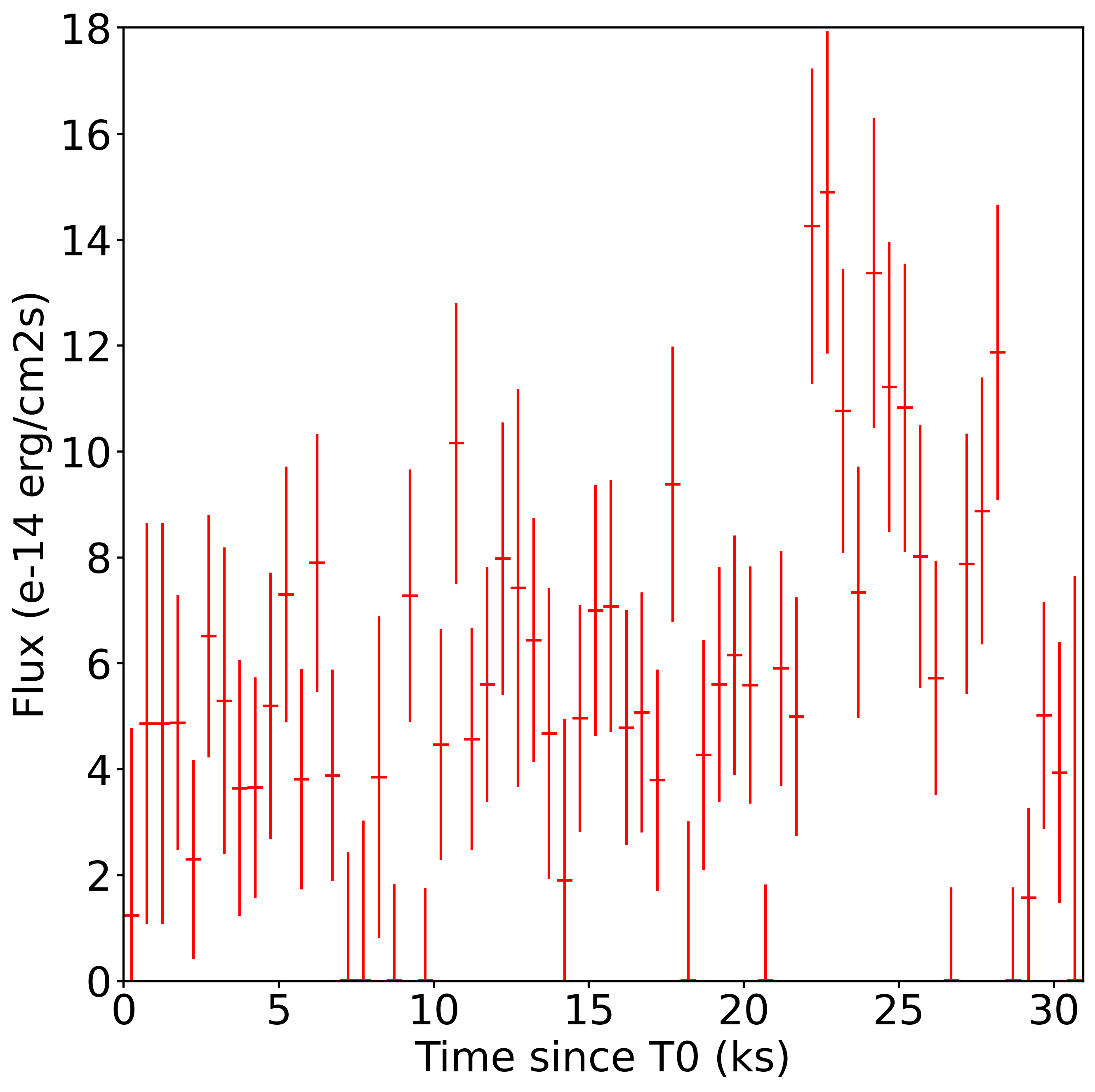}\\
\includegraphics[width=0.8\columnwidth]{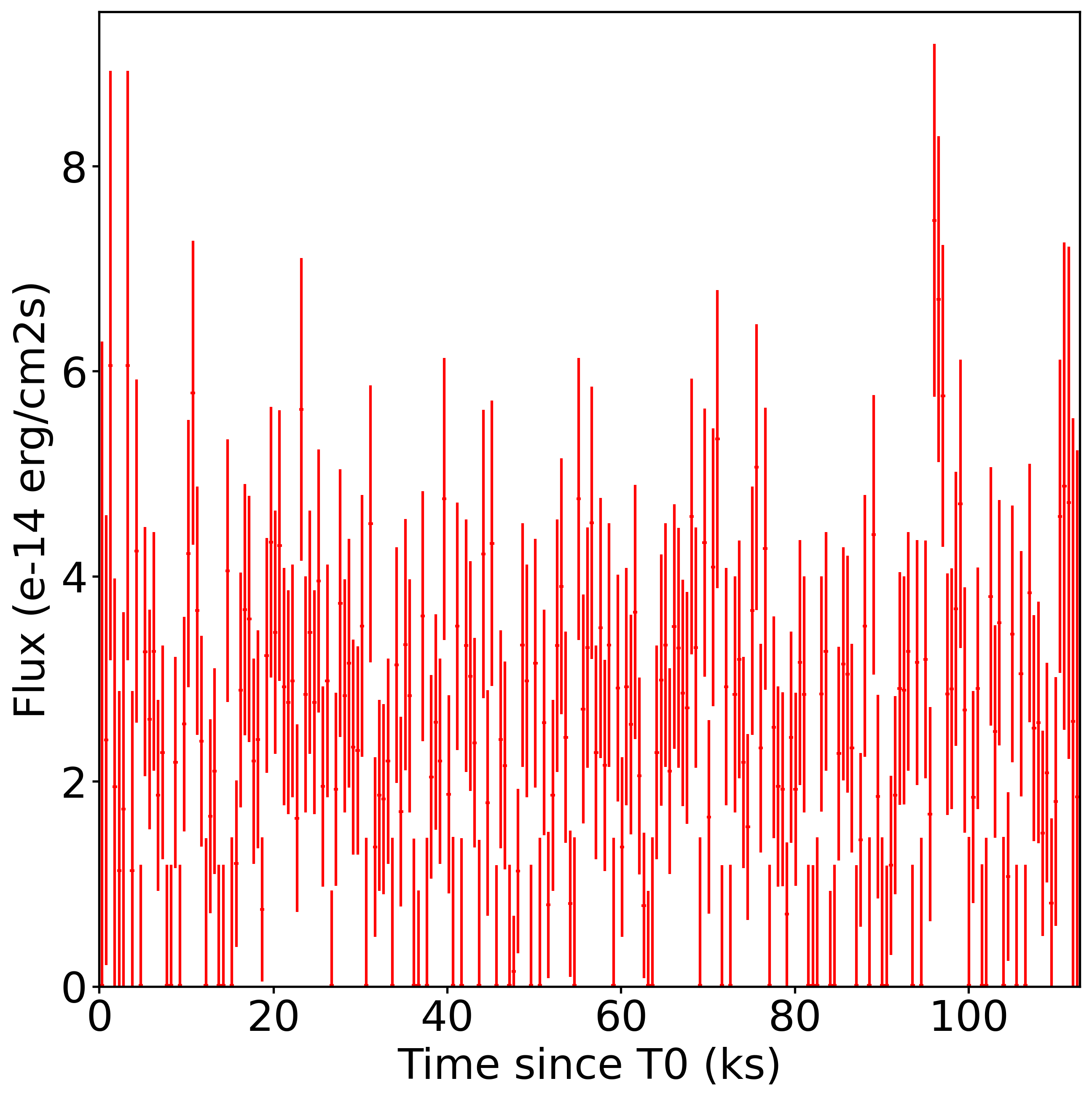}\\
\end{tabular}
\caption{Light curves for false negative sources $0728560301/4$, $0841320101/2$, $0822200101/6$. The variability of the first LC has been ascribed to three random flares; the second LC shows a feature around t$\sim$22 ks; the third LC shows a probable short ($\sim$1 ks) flare at t$\sim$95 ks. T0 is the time of the first photon of the observation. \label{fanecurve}}
\end{figure}

\begin{figure}
\begin{tabular}{c}
\includegraphics[width=\columnwidth]{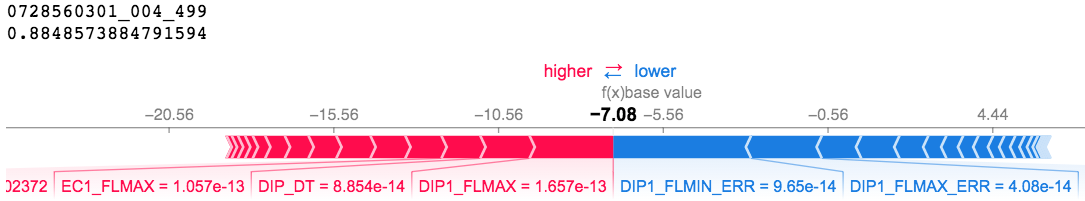} \\
\includegraphics[width=\columnwidth]{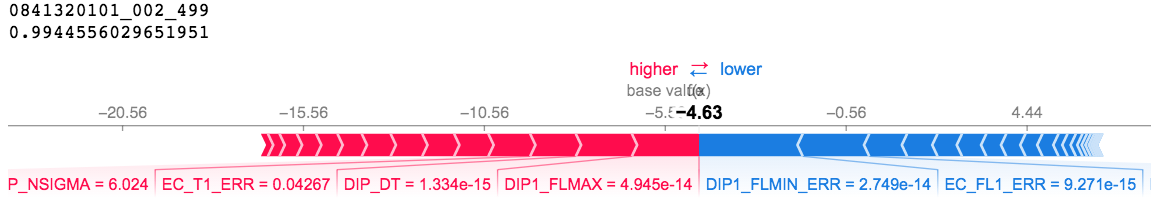} \\
\includegraphics[width=\columnwidth]{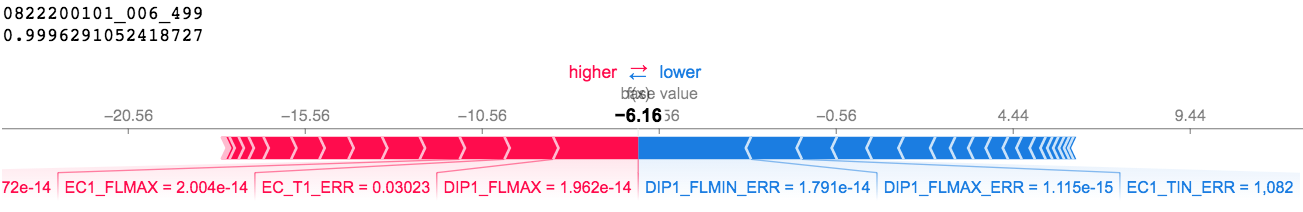} \\
\end{tabular}
\caption{Shapley values for false negative source $0728560301/4$ (top), $0841320101/2$ (mid), $0822200101/6$ (bottom).\label{fane123}}
\end{figure}

We carried out a similar analysis for false negative instances. Fig.~\ref{fanecurve} shows three sources confidently classified as not having a flare while the ground truth class was flare. Clearly noise, non-flaring variability, and the presence of multiple flares (top panel) played a role in the misclassification of these LCs. The Shapley plots are shown in Fig.~\ref{fane123}. Here we see that the contribution to the final decision of the classifier comes mostly from features related to eclipse and dip fits. These are not among the important features we discussed above, but are playing an important role for these sources, possibly because the usual features did not provide a clear cut result.

\subsection{The catalog of candidate flares}

\begin{table}
\caption{First lines of the online material fits file, reporting the catalog of candidate flares. OBS\_ID is the observation identifier of the {\it XMM-Newton} observation and SRC\_NUM the source number; these two parameters define the "name" of the source. Then, we report the fk5 celestial coordinates, in degrees (RA, DEC). The last two colums are the results of our model: ST\_FLARE show if the source is predicted to be a flare, on the basis of the threshold of 0.5 on the resulting probability, PROB.
\label{esfits}}
\scriptsize
\begin{tabular}{llllll}
\hline
\hline
OBS\_ID & SRC\_NUM & RA & DEC & ST\_FLARE & PROB\\
\hline
0000110101 & 1 & 64.926 & 55.999 & F & 0.0006\\
0000110101 & 6 & 64.996 & 56.225 & F & 0.0002\\
0001730201 & 1 & 263.677 & -32.582 & F & 0.02\\
0001730201 & 9 & 263.649 & -32.596 & T & 0.99\\
\hline
\end{tabular}
\end{table}

We applied the trained model to the $31,832$ EXTraS variable LC with uniform time bin (500 s) generated from XMM-Newton data collected between 2000 and 2020. We considered as candidate flares all LC predicted as flares with probability over $0.5$, corresponding to a precision of $82.4$\% and recall of $73.3$\%. We obtain $2088$ candidate flares, of which we expect $1721 \pm 18$ to be actual stellar flares. 
In the online material we report the the probability of "stellar flareness" coming from our model for each of the $31,832$ sources together with their classification as a flaring LC, that is simply based on a threshold of $0.5$ on the probability, as already discussed. Each source is defined by a combination of observation id and source number, as well as the associated celestial coordinates reported in the XMM catalog. Here we note that we run the EXTraS tools using the XMM catalog version available at the moment of the run or, if not available for a given observation we used the standard results from the XMM-Newton Pipeline Processing System (PPS)\footnote{Since minor changes to the detection tools take place across different versions of the Science Analysis System (SAS) and different releases of the catalog not only add new entries but also re-run the detection for some observations, we warn that the reported coordinates can slightly change if compared to a given catalog distribution, as well as the source number (that is ordered based on the predicted count rate, from the highest to the lowest).}. Table \ref{esfits} shows the first lines of the online material fits, as an example.
As reported in Sect. \ref{sec:miscl} among the false positives there is a number of real flares that are not associated to a {\it Gaia} or {\it SIMBAD} optical counterpart, while the main source of false negatives seems to be the presence of a large number of flares in the same LC or, more often, a noisy, non-flaring variability.

\begin{figure}
\includegraphics[width=\columnwidth]{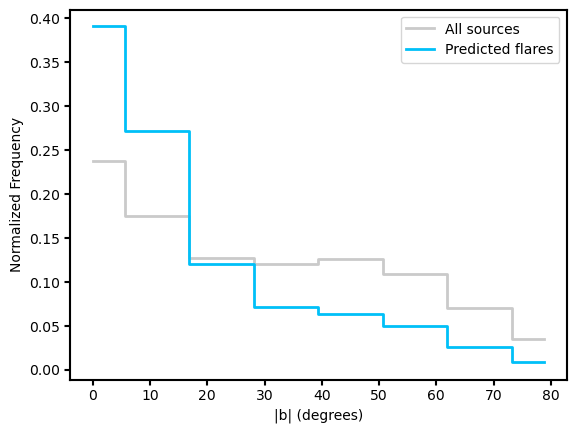}
\caption{Histogram of the modulus of Galactic latitude (in degrees) of sources that have been tagged as flares in the final catalog (cerulean blue) superimposed onto the histogram of all sources in the catalog. The two histograms share the same normalization.\label{flare_b_hist}}
\end{figure}

A simple check that our results are consistent with the expectation based on the stellar origin of flares for the majority of our sources, we show the histogram of the absolute value of Galactic latitude for flaring sources (as defined by our model and threshold) in Fig.~\ref{flare_b_hist}, where we compare it to that of all sources. It is clear from Fig.~\ref{flare_b_hist} that candidate flares are more concentrated towards the Galactic plane, suggesting that indeed they are mostly genuine stellar flares. A two-sample Kolmogorov-Smirnoff test rejects the null hypothesis that candidate flares share the same distribution as the other sources in our catalog with $p = 1.1 \times 10^{-139}$.

\subsection{Bias mitigation}
The final adopted model was not based on any pre-processing or post-processing intended to enforce demographic parity with respect to source brightness or signal-to-noise ratio. We did however explore bias mitigation strategies such as pre-processing to remove correlations with a protected attribute defined as a binary variable BRIGHT corresponding to thresholding AVE against its median. This was carried out using CorrelationRemover in fairlearn on the PFI important features only. It allowed us to reach an approximately equal selection rate (equal fraction of predicted flares among bright sources and dull sources) at the expense of some precision (about $0.4\%$). This is shown in Tab.~\ref{feelthefairness}.

\begin{table}
\caption{Performance of the model trained on PFI-important features compared with an identical model trained on the same features preprocessed with CorrelationRemover. The deviation from demographic parity was measured as the percent difference in the fraction of objects predicted as flares on the test set by either classifier.\label{feelthefairness}}
\begin{tabular}{lllll}
\hline
\hline
Model & Accuracy & Precision & Recall & Deviation from\\
 &  &  &  &dem. parity\\
\hline
Imp. feat. & $96.2\%$ & $77.8\%$ & $70.9\%$ & $13.7\%$\\
Corr. Rem. & $96.2\%$ & $77.4\%$ & $70.9\%$ & $0.07\%$ \\
\hline
\end{tabular}
\end{table}

\section{Discussion and conclusions}
\label{sec:discussion}

We have shown that a gradient boosted classifier trained on a set of $108$ variability features derived from X-ray source light curves is able to classify stellar flares with an accuracy of 97.1\%, a precision of 82.4\% and a recall of 73.3\% on an unseen test set. This is a good result for a classifier that works only on summary features, without direct access to light curves. 

We used our classifier to compile and release to the public the largest unbiased (in terms of sky coverage) catalog of stellar flares to date. Further applications of our work may involve classification of sources from the EXtraS archive that are characterized by less clear variability in terms of goodness of fit with a constant, and the extension of our approach to other datasets, such as those that will be produced by the forthcoming ATHENA \citep[][]{2013arXiv1306.2307N} and AXIS \citep[][]{2019BAAS...51g.107M} missions, as well as for e.g. Chandra or eROSITA. Clearly this is feasible as long as comparable features are computed on the relevant light curves and after checking for significant distribution shifts. This is likely more feasible for a feature-based approach such as ours then for a deep learning model where features are learned automatically, reducing our control on their meaning.

Our results show that even a simple machine learning approach can save a considerable amount of work for human expert annotators, who are nonetheless still likely to be required in a data analysis pipeline. We can quantitatively estimate the impact of our work, given $N$ sources of which $p N$ actual flares, i.e. $p$ is the actual prevalence of flares.

Given a model with recall $r$, it will correctly tag as flares $r p N$ sources: remember that recall is

\begin{equation}
r = \frac{\mathrm{TP}}{\mathrm{TP} + \mathrm{FN}}
\end{equation}

I will also tag as flares an additional $s (1-p) N$ sources, where $s$ is the false positive rate

\begin{equation}
s = \frac{\mathrm{FP}}{\mathrm{FP} + \mathrm{TN}} = \frac{r p (1/P - 1)}{1-p}
\end{equation}

where $P$ is precision,

\begin{equation}
P = \frac{\mathrm{TP}}{\mathrm{TP} + \mathrm{FP}}
\end{equation}

So in total a human expert will have to look at
\begin{equation}
(r p + s (1-p)) N = \frac{rp}{P} N
\end{equation}
sources out of at total of $N$, to catch $r p N$ actual flares.

We can now compare two models with precision $P$ and $P^\prime > P$ in terms of how many more sources a human expert will have to look at when using the worse model with respect to the better model. With the model that has precision $P$, to identify $M$ actual flares the expert will need to look at $M/P$ sources; with the model that has precision $P^\prime$ the expert will need to look at $M/P^\prime$ sources.

For our model $P^\prime = 82.4\%$ or, if we force it to have the same recall as the simple cutoff based on template fitting $P^\prime = 66.0\%$, compared to $P = 33.4\%$ for said method. Even in the latter case, the amount of manual labor saved is roughly a factor two. It should be noted that these benefits are not present if we insist to train our classifier on model-independent features only, suggesting that such features are inadequate -in isolation- for capturing flares. This justifies further work directly on light curves, possibly using deep learning to learn features directly from the data.


However, one of the drawbacks of this method is its opaque nature. To alleviate this issue we applied a variety of explainable AI techniques to our trained classifier. We used permutation feature importance scores to identify the most important features among those used by our classifier, which are shown in Tab.~\ref{pfi}.

Additionally, we visualized the space of these important features using UMAP and obtained ICE curves for each feature. ICE curves visualize how our model prediction on a given data point changes by varying only one feature while keeping the others fixed. ICE curves are mostly non-intersecting for the features we considered, suggesting the lack of interactions between features. This is an important clue for building better models: in particular generalized additive models \citep[GAM; ][]{10.1214/ss/1177013604} may be suited to our problem. GAM are inherently interpretable since they predict their outcomes by modeling the relationship between the response variable and individual predictors as a sum of smooth functions. This is viable only in the context of limited interactions between variables.

We also analyzed the false positive and false negative data points with the help of Shapley values, and visualized their light curves. In hindsight most false positives appear to be issues with the labels: a human expert would, on second thought, classify those as flares as well. Moving forward it seems crucial to develop a standardized protocol for the visual inspection of flares to be used as training instances; this is however beyond the scope of the current paper. False negatives are genuine mistakes of our model, but are anyway often peculiar curves such as multiple flares.

\begin{acknowledgements}
The authors thank the anonymous referee for pointing out several issues in the original version of the manuscript. This work was funded in part by the INAF large grant for the project Astronomy with Natively Interpretable MAchine learning (ANIMA, PI Mario Pasquato).  
\end{acknowledgements}

\section*{Data Availability}

The code used to train the models and the trained models will be released upon publication. 


\begin{thebibliography}{54}
\expandafter\ifx\csname natexlab\endcsname\relax\def\natexlab#1{#1}\fi

\bibitem[{Altmann {et~al.}(2010)Altmann, Tolo{\c{s}}i, Sander, \&
  Lengauer}]{altmann2010permutation}
Altmann, A., Tolo{\c{s}}i, L., Sander, O., \& Lengauer, T. 2010,
  Bioinformatics, 26, 1340

\bibitem[{{Bevington}(1969)}]{1969drea.book.....B}
{Bevington}, P.~R. 1969, {Data reduction and error analysis for the physical
  sciences}

\bibitem[{Bird {et~al.}(2020)Bird, Dud{\'i}k, Edgar, Horn, Lutz, Milan, Sameki,
  Wallach, \& Walker}]{bird2020fairlearn}
Bird, S., Dud{\'i}k, M., Edgar, R., {et~al.} 2020, Fairlearn: A toolkit for
  assessing and improving fairness in {AI}, Tech. Rep. MSR-TR-2020-32,
  Microsoft

\bibitem[{Caton \& Haas(2024)}]{10.1145/3616865}
Caton, S. \& Haas, C. 2024, ACM Comput. Surv., 56

\bibitem[{{Chawla} {et~al.}(2011){Chawla}, {Bowyer}, {Hall}, \&
  {Kegelmeyer}}]{2011arXiv1106.1813C}
{Chawla}, N.~V., {Bowyer}, K.~W., {Hall}, L.~O., \& {Kegelmeyer}, W.~P. 2011,
  arXiv e-prints, arXiv:1106.1813

\bibitem[{{Chen} \& {Guestrin}(2016)}]{2016arXiv160302754C}
{Chen}, T. \& {Guestrin}, C. 2016, arXiv e-prints, arXiv:1603.02754

\bibitem[{{De Luca} {et~al.}(2022){De Luca}, {Israel}, {Salvaterra},
  {Belfiore}, {De Martino}, {Esposito}, {Kovacevic}, {Marelli}, {Mereghetti},
  {Mignani}, {Motta}, {Novara}, {Pasquato}, {Pintore}, {Pizzocaro},
  {Rodriguez-Castillo}, {Sidoli}, {Stelzer}, {Tiengo}, {Wolter}, \&
  {Zampieri}}]{2022MmSAI..93b.122D}
{De Luca}, A., {Israel}, G.~L., {Salvaterra}, R., {et~al.} 2022, in Memorie
  della Societa Astronomica Italiana, Vol.~93, 122

\bibitem[{{De Luca} {et~al.}(2021){De Luca}, {Salvaterra}, {Belfiore},
  {Carpano}, {D'Agostino}, {Haberl}, {Israel}, {Law-Green}, {Lisini},
  {Marelli}, {Novara}, {Read}, {Rodriguez-Castillo}, {Rosen}, {Salvetti},
  {Tiengo}, {Vianello}, {Watson}, {Delvaux}, {Dickens}, {Esposito}, {Greiner},
  {H{\"a}mmerle}, {Kreikenbohm}, {Kreykenbohm}, {Oertel}, {Pizzocaro}, {Pye},
  {Sandrelli}, {Stelzer}, {Wilms}, \& {Zagaria}}]{2021A&A...650A.167D}
{De Luca}, A., {Salvaterra}, R., {Belfiore}, A., {et~al.} 2021, \aap, 650, A167

\bibitem[{{De Luca} {et~al.}(2020){De Luca}, {Stelzer}, {Burgasser},
  {Pizzocaro}, {Ranalli}, {Raetz}, {Marelli}, {Novara}, {Vignali}, {Belfiore},
  {Esposito}, {Franzetti}, {Fumana}, {Gilli}, {Salvaterra}, \&
  {Tiengo}}]{2020A&A...634L..13D}
{De Luca}, A., {Stelzer}, B., {Burgasser}, A.~J., {et~al.} 2020, \aap, 634, L13

\bibitem[{{Diaconis} {et~al.}(2008){Diaconis}, {Goel}, \&
  {Holmes}}]{2008arXiv0811.1477D}
{Diaconis}, P., {Goel}, S., \& {Holmes}, S. 2008, arXiv e-prints,
  arXiv:0811.1477

\bibitem[{{Dillmann} {et~al.}(2025){Dillmann}, {Mart{\'\i}nez-Galarza},
  {Soria}, {Stefano}, \& {Kashyap}}]{2025MNRAS.537..931D}
{Dillmann}, S., {Mart{\'\i}nez-Galarza}, J.~R., {Soria}, R., {Stefano}, R.~D.,
  \& {Kashyap}, V.~L. 2025, \mnras, 537, 931

\bibitem[{{Farrell} {et~al.}(2015){Farrell}, {Murphy}, \&
  {Lo}}]{2015ApJ...813...28F}
{Farrell}, S.~A., {Murphy}, T., \& {Lo}, K.~K. 2015, \apj, 813, 28

\bibitem[{Friedman {et~al.}(2000)Friedman, Hastie, \&
  Tibshirani}]{friedman2000additive}
Friedman, J., Hastie, T., \& Tibshirani, R. 2000, The annals of statistics, 28,
  337

\bibitem[{Friedman(2001)}]{friedman2001greedy}
Friedman, J.~H. 2001, Annals of statistics, 1189

\bibitem[{{Gaia Collaboration} {et~al.}(2016){Gaia Collaboration}, {Prusti},
  {de Bruijne}, {Brown}, {Vallenari}, {Babusiaux}, {Bailer-Jones}, {Bastian},
  {Biermann}, {Evans}, {Eyer}, {Jansen}, {Jordi}, {Klioner}, {Lammers},
  {Lindegren}, {Luri}, {Mignard}, {Milligan}, {Panem}, {Poinsignon},
  {Pourbaix}, {Randich}, {Sarri}, {Sartoretti}, {Siddiqui}, {Soubiran},
  {Valette}, {van Leeuwen}, {Walton}, {Aerts}, {Arenou}, {Cropper}, {Drimmel},
  {H{\o}g}, {Katz}, {Lattanzi}, {O'Mullane}, {Grebel}, {Holland}, {Huc},
  {Passot}, {Bramante}, {Cacciari}, {Casta{\~n}eda}, {Chaoul}, {Cheek}, {De
  Angeli}, {Fabricius}, {Guerra}, {Hern{\'a}ndez}, {Jean-Antoine-Piccolo},
  {Masana}, {Messineo}, {Mowlavi}, {Nienartowicz}, {Ord{\'o}{\~n}ez-Blanco},
  {Panuzzo}, {Portell}, {Richards}, {Riello}, {Seabroke}, {Tanga},
  {Th{\'e}venin}, {Torra}, {Els}, {Gracia-Abril}, {Comoretto},
  {Garcia-Reinaldos}, {Lock}, {Mercier}, {Altmann}, {Andrae}, {Astraatmadja},
  {Bellas-Velidis}, {Benson}, {Berthier}, {Blomme}, {Busso}, {Carry},
  {Cellino}, {Clementini}, {Cowell}, {Creevey}, {Cuypers}, {Davidson}, {De
  Ridder}, {de Torres}, {Delchambre}, {Dell'Oro}, {Ducourant}, {Fr{\'e}mat},
  {Garc{\'\i}a-Torres}, {Gosset}, {Halbwachs}, {Hambly}, {Harrison}, {Hauser},
  {Hestroffer}, {Hodgkin}, {Huckle}, {Hutton}, {Jasniewicz}, {Jordan},
  {Kontizas}, {Korn}, {Lanzafame}, {Manteiga}, {Moitinho}, {Muinonen},
  {Osinde}, {Pancino}, {Pauwels}, {Petit}, {Recio-Blanco}, {Robin}, {Sarro},
  {Siopis}, {Smith}, {Smith}, {Sozzetti}, {Thuillot}, {van Reeven}, {Viala},
  {Abbas}, {Abreu Aramburu}, {Accart}, {Aguado}, {Allan}, {Allasia},
  {Altavilla}, {{\'A}lvarez}, {Alves}, {Anderson}, {Andrei}, {Anglada Varela},
  {Antiche}, {Antoja}, {Ant{\'o}n}, {Arcay}, {Atzei}, {Ayache}, {Bach},
  {Baker}, {Balaguer-N{\'u}{\~n}ez}, {Barache}, {Barata}, {Barbier}, {Barblan},
  {Baroni}, {Barrado y Navascu{\'e}s}, {Barros}, {Barstow}, {Becciani},
  {Bellazzini}, {Bellei}, {Bello Garc{\'\i}a}, {Belokurov}, {Bendjoya},
  {Berihuete}, {Bianchi}, {Bienaym{\'e}}, {Billebaud}, {Blagorodnova},
  {Blanco-Cuaresma}, {Boch}, {Bombrun}, {Borrachero}, {Bouquillon}, {Bourda},
  {Bouy}, {Bragaglia}, {Breddels}, {Brouillet}, {Br{\"u}semeister},
  {Bucciarelli}, {Budnik}, {Burgess}, {Burgon}, {Burlacu}, {Busonero}, {Buzzi},
  {Caffau}, {Cambras}, {Campbell}, {Cancelliere}, {Cantat-Gaudin}, {Carlucci},
  {Carrasco}, {Castellani}, {Charlot}, {Charnas}, {Charvet}, {Chassat},
  {Chiavassa}, {Clotet}, {Cocozza}, {Collins}, {Collins}, {Costigan}, {Crifo},
  {Cross}, {Crosta}, {Crowley}, {Dafonte}, {Damerdji}, {Dapergolas}, {David},
  {David}, {De Cat}, {de Felice}, {de Laverny}, {De Luise}, {De March}, {de
  Martino}, {de Souza}, {Debosscher}, {del Pozo}, {Delbo}, {Delgado},
  {Delgado}, {di Marco}, {Di Matteo}, {Diakite}, {Distefano}, {Dolding}, {Dos
  Anjos}, {Drazinos}, {Dur{\'a}n}, {Dzigan}, {Ecale}, {Edvardsson}, {Enke},
  {Erdmann}, {Escolar}, {Espina}, {Evans}, {Eynard Bontemps}, {Fabre},
  {Fabrizio}, {Faigler}, {Falc{\~a}o}, {Farr{\`a}s Casas}, {Faye}, {Federici},
  {Fedorets}, {Fern{\'a}ndez-Hern{\'a}ndez}, {Fernique}, {Fienga}, {Figueras},
  {Filippi}, {Findeisen}, {Fonti}, {Fouesneau}, {Fraile}, {Fraser}, {Fuchs},
  {Furnell}, {Gai}, {Galleti}, {Galluccio}, {Garabato}, {Garc{\'\i}a-Sedano},
  {Gar{\'e}}, {Garofalo}, {Garralda}, {Gavras}, {Gerssen}, {Geyer}, {Gilmore},
  {Girona}, {Giuffrida}, {Gomes}, {Gonz{\'a}lez-Marcos},
  {Gonz{\'a}lez-N{\'u}{\~n}ez}, {Gonz{\'a}lez-Vidal}, {Granvik}, {Guerrier},
  {Guillout}, {Guiraud}, {G{\'u}rpide}, {Guti{\'e}rrez-S{\'a}nchez}, {Guy},
  {Haigron}, {Hatzidimitriou}, {Haywood}, {Heiter}, {Helmi}, {Hobbs},
  {Hofmann}, {Holl}, {Holland}, {Hunt}, {Hypki}, {Icardi}, {Irwin}, {Jevardat
  de Fombelle}, {Jofr{\'e}}, {Jonker}, {Jorissen}, {Julbe}, {Karampelas},
  {Kochoska}, {Kohley}, {Kolenberg}, {Kontizas}, {Koposov}, {Kordopatis},
  {Koubsky}, {Kowalczyk}, {Krone-Martins}, {Kudryashova}, {Kull}, {Bachchan},
  {Lacoste-Seris}, {Lanza}, {Lavigne}, {Le Poncin-Lafitte}, {Lebreton},
  {Lebzelter}, {Leccia}, {Leclerc}, {Lecoeur-Taibi}, {Lemaitre}, {Lenhardt},
  {Leroux}, {Liao}, {Licata}, {Lindstr{\o}m}, {Lister}, {Livanou}, {Lobel},
  {L{\"o}ffler}, {L{\'o}pez}, {Lopez-Lozano}, {Lorenz}, {Loureiro},
  {MacDonald}, {Magalh{\~a}es Fernandes}, {Managau}, {Mann}, {Mantelet},
  {Marchal}, {Marchant}, {Marconi}, {Marie}, {Marinoni}, {Marrese},
  {Marschalk{\'o}}, {Marshall}, {Mart{\'\i}n-Fleitas}, {Martino}, {Mary},
  {Matijevi{\v{c}}}, {Mazeh}, {McMillan}, {Messina}, {Mestre}, {Michalik},
  {Millar}, {Miranda}, {Molina}, {Molinaro}, {Molinaro}, {Moln{\'a}r},
  {Moniez}, {Montegriffo}, {Monteiro}, {Mor}, {Mora}, {Morbidelli}, {Morel},
  {Morgenthaler}, {Morley}, {Morris}, {Mulone}, {Muraveva}, {Musella},
  {Narbonne}, {Nelemans}, {Nicastro}, {Noval}, {Ord{\'e}novic},
  {Ordieres-Mer{\'e}}, {Osborne}, {Pagani}, {Pagano}, {Pailler}, {Palacin},
  {Palaversa}, {Parsons}, {Paulsen}, {Pecoraro}, {Pedrosa}, {Pentik{\"a}inen},
  {Pereira}, {Pichon}, {Piersimoni}, {Pineau}, {Plachy}, {Plum}, {Poujoulet},
  {Pr{\v{s}}a}, {Pulone}, {Ragaini}, {Rago}, {Rambaux}, {Ramos-Lerate},
  {Ranalli}, {Rauw}, {Read}, {Regibo}, {Renk}, {Reyl{\'e}}, {Ribeiro},
  {Rimoldini}, {Ripepi}, {Riva}, {Rixon}, {Roelens}, {Romero-G{\'o}mez},
  {Rowell}, {Royer}, {Rudolph}, {Ruiz-Dern}, {Sadowski}, {Sagrist{\`a}
  Sell{\'e}s}, {Sahlmann}, {Salgado}, {Salguero}, {Sarasso}, {Savietto},
  {Schnorhk}, {Schultheis}, {Sciacca}, {Segol}, {Segovia}, {Segransan},
  {Serpell}, {Shih}, {Smareglia}, {Smart}, {Smith}, {Solano}, {Solitro},
  {Sordo}, {Soria Nieto}, {Souchay}, {Spagna}, {Spoto}, {Stampa}, {Steele},
  {Steidelm{\"u}ller}, {Stephenson}, {Stoev}, {Suess}, {S{\"u}veges}, {Surdej},
  {Szabados}, {Szegedi-Elek}, {Tapiador}, {Taris}, {Tauran}, {Taylor},
  {Teixeira}, {Terrett}, {Tingley}, {Trager}, {Turon}, {Ulla}, {Utrilla},
  {Valentini}, {van Elteren}, {Van Hemelryck}, {van Leeuwen}, {Varadi},
  {Vecchiato}, {Veljanoski}, {Via}, {Vicente}, {Vogt}, {Voss}, {Votruba},
  {Voutsinas}, {Walmsley}, {Weiler}, {Weingrill}, {Werner}, {Wevers},
  {Whitehead}, {Wyrzykowski}, {Yoldas}, {{\v{Z}}erjal}, {Zucker}, {Zurbach},
  {Zwitter}, {Alecu}, {Allen}, {Allende Prieto}, {Amorim},
  {Anglada-Escud{\'e}}, {Arsenijevic}, {Azaz}, {Balm}, {Beck}, {Bernstein},
  {Bigot}, {Bijaoui}, {Blasco}, {Bonfigli}, {Bono}, {Boudreault}, {Bressan},
  {Brown}, {Brunet}, {Bunclark}, {Buonanno}, {Butkevich}, {Carret}, {Carrion},
  {Chemin}, {Ch{\'e}reau}, {Corcione}, {Darmigny}, {de Boer}, {de Teodoro}, {de
  Zeeuw}, {Delle Luche}, {Domingues}, {Dubath}, {Fodor}, {Fr{\'e}zouls},
  {Fries}, {Fustes}, {Fyfe}, {Gallardo}, {Gallegos}, {Gardiol}, {Gebran},
  {Gomboc}, {G{\'o}mez}, {Grux}, {Gueguen}, {Heyrovsky}, {Hoar}, {Iannicola},
  {Isasi Parache}, {Janotto}, {Joliet}, {Jonckheere}, {Keil}, {Kim},
  {Klagyivik}, {Klar}, {Knude}, {Kochukhov}, {Kolka}, {Kos}, {Kutka}, {Lainey},
  {LeBouquin}, {Liu}, {Loreggia}, {Makarov}, {Marseille}, {Martayan},
  {Martinez-Rubi}, {Massart}, {Meynadier}, {Mignot}, {Munari}, {Nguyen},
  {Nordlander}, {Ocvirk}, {O'Flaherty}, {Olias Sanz}, {Ortiz}, {Osorio},
  {Oszkiewicz}, {Ouzounis}, {Palmer}, {Park}, {Pasquato}, {Peltzer}, {Peralta},
  {P{\'e}turaud}, {Pieniluoma}, {Pigozzi}, {Poels}, {Prat}, {Prod'homme},
  {Raison}, {Rebordao}, {Risquez}, {Rocca-Volmerange}, {Rosen}, {Ruiz-Fuertes},
  {Russo}, {Sembay}, {Serraller Vizcaino}, {Short}, {Siebert}, {Silva},
  {Sinachopoulos}, {Slezak}, {Soffel}, {Sosnowska}, {Strai{\v{z}}ys}, {ter
  Linden}, {Terrell}, {Theil}, {Tiede}, {Troisi}, {Tsalmantza}, {Tur},
  {Vaccari}, {Vachier}, {Valles}, {Van Hamme}, {Veltz}, {Virtanen}, {Wallut},
  {Wichmann}, {Wilkinson}, {Ziaeepour}, \& {Zschocke}}]{2016A&A...595A...1G}
{Gaia Collaboration}, {Prusti}, T., {de Bruijne}, J.~H.~J., {et~al.} 2016,
  \aap, 595, A1

\bibitem[{{Gaia Collaboration} {et~al.}(2023){Gaia Collaboration}, {Vallenari},
  {Brown}, {Prusti}, {de Bruijne}, {Arenou}, {Babusiaux}, {Biermann},
  {Creevey}, {Ducourant}, {Evans}, {Eyer}, {Guerra}, {Hutton}, {Jordi},
  {Klioner}, {Lammers}, {Lindegren}, {Luri}, {Mignard}, {Panem}, {Pourbaix},
  {Randich}, {Sartoretti}, {Soubiran}, {Tanga}, {Walton}, {Bailer-Jones},
  {Bastian}, {Drimmel}, {Jansen}, {Katz}, {Lattanzi}, {van Leeuwen}, {Bakker},
  {Cacciari}, {Casta{\~n}eda}, {De Angeli}, {Fabricius}, {Fouesneau},
  {Fr{\'e}mat}, {Galluccio}, {Guerrier}, {Heiter}, {Masana}, {Messineo},
  {Mowlavi}, {Nicolas}, {Nienartowicz}, {Pailler}, {Panuzzo}, {Riclet}, {Roux},
  {Seabroke}, {Sordo}, {Th{\'e}venin}, {Gracia-Abril}, {Portell}, {Teyssier},
  {Altmann}, {Andrae}, {Audard}, {Bellas-Velidis}, {Benson}, {Berthier},
  {Blomme}, {Burgess}, {Busonero}, {Busso}, {C{\'a}novas}, {Carry}, {Cellino},
  {Cheek}, {Clementini}, {Damerdji}, {Davidson}, {de Teodoro}, {Nu{\~n}ez
  Campos}, {Delchambre}, {Dell'Oro}, {Esquej}, {Fern{\'a}ndez-Hern{\'a}ndez},
  {Fraile}, {Garabato}, {Garc{\'\i}a-Lario}, {Gosset}, {Haigron}, {Halbwachs},
  {Hambly}, {Harrison}, {Hern{\'a}ndez}, {Hestroffer}, {Hodgkin}, {Holl},
  {Jan{\ss}en}, {Jevardat de Fombelle}, {Jordan}, {Krone-Martins}, {Lanzafame},
  {L{\"o}ffler}, {Marchal}, {Marrese}, {Moitinho}, {Muinonen}, {Osborne},
  {Pancino}, {Pauwels}, {Recio-Blanco}, {Reyl{\'e}}, {Riello}, {Rimoldini},
  {Roegiers}, {Rybizki}, {Sarro}, {Siopis}, {Smith}, {Sozzetti}, {Utrilla},
  {van Leeuwen}, {Abbas}, {{\'A}brah{\'a}m}, {Abreu Aramburu}, {Aerts},
  {Aguado}, {Ajaj}, {Aldea-Montero}, {Altavilla}, {{\'A}lvarez}, {Alves},
  {Anders}, {Anderson}, {Anglada Varela}, {Antoja}, {Baines}, {Baker},
  {Balaguer-N{\'u}{\~n}ez}, {Balbinot}, {Balog}, {Barache}, {Barbato},
  {Barros}, {Barstow}, {Bartolom{\'e}}, {Bassilana}, {Bauchet}, {Becciani},
  {Bellazzini}, {Berihuete}, {Bernet}, {Bertone}, {Bianchi}, {Binnenfeld},
  {Blanco-Cuaresma}, {Blazere}, {Boch}, {Bombrun}, {Bossini}, {Bouquillon},
  {Bragaglia}, {Bramante}, {Breedt}, {Bressan}, {Brouillet}, {Brugaletta},
  {Bucciarelli}, {Burlacu}, {Butkevich}, {Buzzi}, {Caffau}, {Cancelliere},
  {Cantat-Gaudin}, {Carballo}, {Carlucci}, {Carnerero}, {Carrasco},
  {Casamiquela}, {Castellani}, {Castro-Ginard}, {Chaoul}, {Charlot}, {Chemin},
  {Chiaramida}, {Chiavassa}, {Chornay}, {Comoretto}, {Contursi}, {Cooper},
  {Cornez}, {Cowell}, {Crifo}, {Cropper}, {Crosta}, {Crowley}, {Dafonte},
  {Dapergolas}, {David}, {David}, {de Laverny}, {De Luise}, {De March}, {De
  Ridder}, {de Souza}, {de Torres}, {del Peloso}, {del Pozo}, {Delbo},
  {Delgado}, {Delisle}, {Demouchy}, {Dharmawardena}, {Di Matteo}, {Diakite},
  {Diener}, {Distefano}, {Dolding}, {Edvardsson}, {Enke}, {Fabre}, {Fabrizio},
  {Faigler}, {Fedorets}, {Fernique}, {Fienga}, {Figueras}, {Fournier},
  {Fouron}, {Fragkoudi}, {Gai}, {Garcia-Gutierrez}, {Garcia-Reinaldos},
  {Garc{\'\i}a-Torres}, {Garofalo}, {Gavel}, {Gavras}, {Gerlach}, {Geyer},
  {Giacobbe}, {Gilmore}, {Girona}, {Giuffrida}, {Gomel}, {Gomez},
  {Gonz{\'a}lez-N{\'u}{\~n}ez}, {Gonz{\'a}lez-Santamar{\'\i}a},
  {Gonz{\'a}lez-Vidal}, {Granvik}, {Guillout}, {Guiraud},
  {Guti{\'e}rrez-S{\'a}nchez}, {Guy}, {Hatzidimitriou}, {Hauser}, {Haywood},
  {Helmer}, {Helmi}, {Sarmiento}, {Hidalgo}, {Hilger}, {H{\l}adczuk}, {Hobbs},
  {Holland}, {Huckle}, {Jardine}, {Jasniewicz}, {Jean-Antoine Piccolo},
  {Jim{\'e}nez-Arranz}, {Jorissen}, {Juaristi Campillo}, {Julbe}, {Karbevska},
  {Kervella}, {Khanna}, {Kontizas}, {Kordopatis}, {Korn}, {K{\'o}sp{\'a}l},
  {Kostrzewa-Rutkowska}, {Kruszy{\'n}ska}, {Kun}, {Laizeau}, {Lambert},
  {Lanza}, {Lasne}, {Le Campion}, {Lebreton}, {Lebzelter}, {Leccia}, {Leclerc},
  {Lecoeur-Taibi}, {Liao}, {Licata}, {Lindstr{\o}m}, {Lister}, {Livanou},
  {Lobel}, {Lorca}, {Loup}, {Madrero Pardo}, {Magdaleno Romeo}, {Managau},
  {Mann}, {Manteiga}, {Marchant}, {Marconi}, {Marcos}, {Marcos Santos},
  {Mar{\'\i}n Pina}, {Marinoni}, {Marocco}, {Marshall}, {Martin Polo},
  {Mart{\'\i}n-Fleitas}, {Marton}, {Mary}, {Masip}, {Massari},
  {Mastrobuono-Battisti}, {Mazeh}, {McMillan}, {Messina}, {Michalik}, {Millar},
  {Mints}, {Molina}, {Molinaro}, {Moln{\'a}r}, {Monari}, {Mongui{\'o}},
  {Montegriffo}, {Montero}, {Mor}, {Mora}, {Morbidelli}, {Morel}, {Morris},
  {Muraveva}, {Murphy}, {Musella}, {Nagy}, {Noval}, {Oca{\~n}a}, {Ogden},
  {Ordenovic}, {Osinde}, {Pagani}, {Pagano}, {Palaversa}, {Palicio},
  {Pallas-Quintela}, {Panahi}, {Payne-Wardenaar}, {Pe{\~n}alosa Esteller},
  {Penttil{\"a}}, {Pichon}, {Piersimoni}, {Pineau}, {Plachy}, {Plum}, {Poggio},
  {Pr{\v{s}}a}, {Pulone}, {Racero}, {Ragaini}, {Rainer}, {Raiteri}, {Rambaux},
  {Ramos}, {Ramos-Lerate}, {Re Fiorentin}, {Regibo}, {Richards}, {Rios Diaz},
  {Ripepi}, {Riva}, {Rix}, {Rixon}, {Robichon}, {Robin}, {Robin}, {Roelens},
  {Rogues}, {Rohrbasser}, {Romero-G{\'o}mez}, {Rowell}, {Royer}, {Ruz Mieres},
  {Rybicki}, {Sadowski}, {S{\'a}ez N{\'u}{\~n}ez}, {Sagrist{\`a} Sell{\'e}s},
  {Sahlmann}, {Salguero}, {Samaras}, {Sanchez Gimenez}, {Sanna},
  {Santove{\~n}a}, {Sarasso}, {Schultheis}, {Sciacca}, {Segol}, {Segovia},
  {S{\'e}gransan}, {Semeux}, {Shahaf}, {Siddiqui}, {Siebert}, {Siltala},
  {Silvelo}, {Slezak}, {Slezak}, {Smart}, {Snaith}, {Solano}, {Solitro},
  {Souami}, {Souchay}, {Spagna}, {Spina}, {Spoto}, {Steele},
  {Steidelm{\"u}ller}, {Stephenson}, {S{\"u}veges}, {Surdej}, {Szabados},
  {Szegedi-Elek}, {Taris}, {Taylor}, {Teixeira}, {Tolomei}, {Tonello}, {Torra},
  {Torra}, {Torralba Elipe}, {Trabucchi}, {Tsounis}, {Turon}, {Ulla}, {Unger},
  {Vaillant}, {van Dillen}, {van Reeven}, {Vanel}, {Vecchiato}, {Viala},
  {Vicente}, {Voutsinas}, {Weiler}, {Wevers}, {Wyrzykowski}, {Yoldas}, {Yvard},
  {Zhao}, {Zorec}, {Zucker}, \& {Zwitter}}]{2023A&A...674A...1G}
{Gaia Collaboration}, {Vallenari}, A., {Brown}, A.~G.~A., {et~al.} 2023, \aap,
  674, A1

\bibitem[{Gunning {et~al.}(2019)Gunning, Stefik, Choi, Miller, Stumpf, \&
  Yang}]{gunning2019xai}
Gunning, D., Stefik, M., Choi, J., {et~al.} 2019, Science robotics, 4, eaay7120

\bibitem[{Hastie \& Tibshirani(1986)}]{10.1214/ss/1177013604}
Hastie, T. \& Tibshirani, R. 1986, Statistical Science, 1, 297

\bibitem[{{Huppenkothen} {et~al.}(2023){Huppenkothen}, {Ntampaka}, {Ho},
  {Fouesneau}, {Nord}, {Peek}, {Walmsley}, {Wu}, {Avestruz}, {Buck}, {Brescia},
  {Finkbeiner}, {Goulding}, {Kacprzak}, {Melchior}, {Pasquato}, {Ramachandra},
  {Ting}, {van de Ven}, {Villar}, {Villar}, \& {Zinger}}]{2023arXiv231012528H}
{Huppenkothen}, D., {Ntampaka}, M., {Ho}, M., {et~al.} 2023, arXiv e-prints,
  arXiv:2310.12528

\bibitem[{{Kova{\v{c}}evi{\'c}} {et~al.}(2022){Kova{\v{c}}evi{\'c}},
  {Pasquato}, {Marelli}, {De Luca}, {Salvaterra}, \&
  {Belfiore}}]{2022A&A...659A..66K}
{Kova{\v{c}}evi{\'c}}, M., {Pasquato}, M., {Marelli}, M., {et~al.} 2022, \aap,
  659, A66

\bibitem[{{Kowalski}(2024)}]{2024LRSP...21....1K}
{Kowalski}, A.~F. 2024, Living Reviews in Solar Physics, 21, 1

\bibitem[{{Lemaitre} {et~al.}(2016){Lemaitre}, {Nogueira}, \&
  {Aridas}}]{2016arXiv160906570L}
{Lemaitre}, G., {Nogueira}, F., \& {Aridas}, C.~K. 2016, arXiv e-prints,
  arXiv:1609.06570

\bibitem[{{Lin} {et~al.}(2012){Lin}, {Webb}, \& {Barret}}]{2012ApJ...756...27L}
{Lin}, D., {Webb}, N.~A., \& {Barret}, D. 2012, \apj, 756, 27

\bibitem[{{Lo} {et~al.}(2014){Lo}, {Farrell}, {Murphy}, \&
  {Gaensler}}]{2014ApJ...786...20L}
{Lo}, K.~K., {Farrell}, S., {Murphy}, T., \& {Gaensler}, B.~M. 2014, \apj, 786,
  20

\bibitem[{Lundberg \& Lee(2017)}]{lundberg2017unified}
Lundberg, S.~M. \& Lee, S.-I. 2017, Advances in neural information processing
  systems, 30

\bibitem[{{Marelli} {et~al.}(2018){Marelli}, {De Martino}, {Mereghetti}, {De
  Luca}, {Salvaterra}, {Sidoli}, {Israel}, \&
  {Rodriguez}}]{2018ApJ...866..125M}
{Marelli}, M., {De Martino}, D., {Mereghetti}, S., {et~al.} 2018, \apj, 866,
  125

\bibitem[{{Marelli} {et~al.}(2017){Marelli}, {Tiengo}, {De Luca}, {Salvetti},
  {Saronni}, {Sidoli}, {Paizis}, {Salvaterra}, {Belfiore}, {Israel}, {Haberl},
  \& {D'Agostino}}]{2017ApJ...851L..27M}
{Marelli}, M., {Tiengo}, A., {De Luca}, A., {et~al.} 2017, \apjl, 851, L27

\bibitem[{{McInnes} {et~al.}(2018){McInnes}, {Healy}, \&
  {Melville}}]{2018arXiv180203426M}
{McInnes}, L., {Healy}, J., \& {Melville}, J. 2018, arXiv e-prints,
  arXiv:1802.03426

\bibitem[{{Mereghetti} {et~al.}(2018){Mereghetti}, {De Luca}, {Salvetti},
  {Belfiore}, {Marelli}, {Paizis}, {Rigoselli}, {Salvaterra}, {Sidoli}, \&
  {Tiengo}}]{2018A&A...616A..36M}
{Mereghetti}, S., {De Luca}, A., {Salvetti}, D., {et~al.} 2018, \aap, 616, A36

\bibitem[{{Mushotzky} {et~al.}(2019){Mushotzky}, {Aird}, {Barger},
  {Cappelluti}, {Chartas}, {Corrales}, {Eufrasio}, {Fabian}, {Falcone},
  {Gallo}, {Gilli}, {Grant}, {Hardcastle}, {Hodges-Kluck}, {Kara}, {Koss},
  {Li}, {Lisse}, {Loewenstein}, {Markevitch}, {Meyer}, {Miller}, {Mulchaey},
  {Petre}, {Ptak}, {Reynolds}, {Russell}, {Safi-Harb}, {Smith}, {Snios},
  {Tombesi}, {Valencic}, {Walker}, {Williams}, {Winter}, {Yamaguchi}, {Zhang},
  {Arenberg}, {Brandt}, {Burrows}, {Georganopoulos}, {Miller}, {Norman}, \&
  {Rosati}}]{2019BAAS...51g.107M}
{Mushotzky}, R., {Aird}, J., {Barger}, A.~J., {et~al.} 2019, in Bulletin of the
  American Astronomical Society, Vol.~51, 107

\bibitem[{{Nandra} {et~al.}(2013){Nandra}, {Barret}, {Barcons}, {Fabian}, {den
  Herder}, {Piro}, {Watson}, {Adami}, {Aird}, {Afonso}, {Alexander},
  {Argiroffi}, {Amati}, {Arnaud}, {Atteia}, {Audard}, {Badenes}, {Ballet},
  {Ballo}, {Bamba}, {Bhardwaj}, {Stefano Battistelli}, {Becker}, {De Becker},
  {Behar}, {Bianchi}, {Biffi}, {B{\^\i}rzan}, {Bocchino}, {Bogdanov}, {Boirin},
  {Boller}, {Borgani}, {Borm}, {Bouch{\'e}}, {Bourdin}, {Bower}, {Braito},
  {Branchini}, {Branduardi-Raymont}, {Bregman}, {Brenneman}, {Brightman},
  {Br{\"u}ggen}, {Buchner}, {Bulbul}, {Brusa}, {Bursa}, {Caccianiga},
  {Cackett}, {Campana}, {Cappelluti}, {Cappi}, {Carrera}, {Ceballos},
  {Christensen}, {Chu}, {Churazov}, {Clerc}, {Corbel}, {Corral}, {Comastri},
  {Costantini}, {Croston}, {Dadina}, {D'Ai}, {Decourchelle}, {Della Ceca},
  {Dennerl}, {Dolag}, {Done}, {Dovciak}, {Drake}, {Eckert}, {Edge}, {Ettori},
  {Ezoe}, {Feigelson}, {Fender}, {Feruglio}, {Finoguenov}, {Fiore}, {Galeazzi},
  {Gallagher}, {Gandhi}, {Gaspari}, {Gastaldello}, {Georgakakis},
  {Georgantopoulos}, {Gilfanov}, {Gitti}, {Gladstone}, {Goosmann}, {Gosset},
  {Grosso}, {Guedel}, {Guerrero}, {Haberl}, {Hardcastle}, {Heinz}, {Alonso
  Herrero}, {Herv{\'e}}, {Holmstrom}, {Iwasawa}, {Jonker}, {Kaastra}, {Kara},
  {Karas}, {Kastner}, {King}, {Kosenko}, {Koutroumpa}, {Kraft}, {Kreykenbohm},
  {Lallement}, {Lanzuisi}, {Lee}, {Lemoine-Goumard}, {Lobban}, {Lodato},
  {Lovisari}, {Lotti}, {McCharthy}, {McNamara}, {Maggio}, {Maiolino}, {De
  Marco}, {de Martino}, {Mateos}, {Matt}, {Maughan}, {Mazzotta}, {Mendez},
  {Merloni}, {Micela}, {Miceli}, {Mignani}, {Miller}, {Miniutti}, {Molendi},
  {Montez}, {Moretti}, {Motch}, {Naz{\'e}}, {Nevalainen}, {Nicastro}, {Nulsen},
  {Ohashi}, {O'Brien}, {Osborne}, {Oskinova}, {Pacaud}, {Paerels}, {Page},
  {Papadakis}, {Pareschi}, {Petre}, {Petrucci}, {Piconcelli}, {Pillitteri},
  {Pinto}, {de Plaa}, {Pointecouteau}, {Ponman}, {Ponti}, {Porquet}, {Pounds},
  {Pratt}, {Predehl}, {Proga}, {Psaltis}, {Rafferty}, {Ramos-Ceja}, {Ranalli},
  {Rasia}, {Rau}, {Rauw}, {Rea}, {Read}, {Reeves}, {Reiprich}, {Renaud},
  {Reynolds}, {Risaliti}, {Rodriguez}, {Rodriguez Hidalgo}, {Roncarelli},
  {Rosario}, {Rossetti}, {Rozanska}, {Rovilos}, {Salvaterra}, {Salvato}, {Di
  Salvo}, {Sanders}, {Sanz-Forcada}, {Schawinski}, {Schaye}, {Schwope}, \&
  {Sciortino}}]{2013arXiv1306.2307N}
{Nandra}, K., {Barret}, D., {Barcons}, X., {et~al.} 2013, arXiv e-prints,
  arXiv:1306.2307

\bibitem[{{Orwat-Kapola} {et~al.}(2022){Orwat-Kapola}, {Bird}, {Hill},
  {Altamirano}, \& {Huppenkothen}}]{2022MNRAS.509.1269O}
{Orwat-Kapola}, J.~K., {Bird}, A.~J., {Hill}, A.~B., {Altamirano}, D., \&
  {Huppenkothen}, D. 2022, \mnras, 509, 1269

\bibitem[{{Pasquato} {et~al.}(2024){Pasquato}, {Trevisan}, {Askar}, {Lemos},
  {Carenini}, {Mapelli}, \& {Hezaveh}}]{2024ApJ...965...89P}
{Pasquato}, M., {Trevisan}, P., {Askar}, A., {et~al.} 2024, \apj, 965, 89

\bibitem[{Pedregosa {et~al.}(2011)Pedregosa, Varoquaux, Gramfort, Michel,
  Thirion, Grisel, Blondel, Prettenhofer, Weiss, Dubourg, Vanderplas, Passos,
  Cournapeau, Brucher, Perrot, \& Duchesnay}]{scikit-learn}
Pedregosa, F., Varoquaux, G., Gramfort, A., {et~al.} 2011, Journal of Machine
  Learning Research, 12, 2825

\bibitem[{{P{\'e}rez-D{\'\i}az} {et~al.}(2024){P{\'e}rez-D{\'\i}az},
  {Mart{\'\i}nez-Galarza}, {Caicedo}, \& {D'Abrusco}}]{2024MNRAS.528.4852P}
{P{\'e}rez-D{\'\i}az}, V.~S., {Mart{\'\i}nez-Galarza}, J.~R., {Caicedo}, A., \&
  {D'Abrusco}, R. 2024, \mnras, 528, 4852

\bibitem[{{Pizzocaro} {et~al.}(2016){Pizzocaro}, {Stelzer}, {Paladini},
  {Tiengo}, {Lisini}, {Novara}, {Vianello}, {Belfiore}, {Marelli}, {Salvetti},
  {Pillitteri}, {Sciortino}, {D'Agostino}, {Haberl}, {Watson}, {Wilms},
  {Salvaterra}, \& {De Luca}}]{2016A&A...587A..36P}
{Pizzocaro}, D., {Stelzer}, B., {Paladini}, R., {et~al.} 2016, \aap, 587, A36

\bibitem[{{Pye} {et~al.}(2015){Pye}, {Rosen}, {Fyfe}, \&
  {Schr{\"o}der}}]{2015A&A...581A..28P}
{Pye}, J.~P., {Rosen}, S., {Fyfe}, D., \& {Schr{\"o}der}, A.~C. 2015, \aap,
  581, A28

\bibitem[{{Quirola-V{\'a}squez} {et~al.}(2023){Quirola-V{\'a}squez}, {Bauer},
  {Jonker}, {Brandt}, {Yang}, {Levan}, {Xue}, {Eappachen}, {Camacho},
  {Ravasio}, {Zheng}, \& {Luo}}]{2023A&A...675A..44Q}
{Quirola-V{\'a}squez}, J., {Bauer}, F.~E., {Jonker}, P.~G., {et~al.} 2023,
  \aap, 675, A44

\bibitem[{{Quirola-V{\'a}squez} {et~al.}(2022){Quirola-V{\'a}squez}, {Bauer},
  {Jonker}, {Brandt}, {Yang}, {Levan}, {Xue}, {Eappachen}, {Zheng}, \&
  {Luo}}]{2022A&A...663A.168Q}
{Quirola-V{\'a}squez}, J., {Bauer}, F.~E., {Jonker}, P.~G., {et~al.} 2022,
  \aap, 663, A168

\bibitem[{{Richards} {et~al.}(2011){Richards}, {Starr}, {Butler}, {Bloom},
  {Brewer}, {Crellin-Quick}, {Higgins}, {Kennedy}, \&
  {Rischard}}]{2011ApJ...733...10R}
{Richards}, J.~W., {Starr}, D.~L., {Butler}, N.~R., {et~al.} 2011, \apj, 733,
  10

\bibitem[{{Ricketts} {et~al.}(2023){Ricketts}, {Steiner}, {Garraffo},
  {Remillard}, \& {Huppenkothen}}]{2023MNRAS.523.1946R}
{Ricketts}, B.~J., {Steiner}, J.~F., {Garraffo}, C., {Remillard}, R.~A., \&
  {Huppenkothen}, D. 2023, \mnras, 523, 1946

\bibitem[{Rijsbergen(1979)}]{rijsbergen1979information}
Rijsbergen, C.~v. 1979, Information retrieval (Butterworth-Heinemann)

\bibitem[{{Ruiz} {et~al.}(2024){Ruiz}, {Georgakakis}, {Georgantopoulos},
  {Akylas}, {Pierre}, \& {Starck}}]{2024MNRAS.527.3674R}
{Ruiz}, A., {Georgakakis}, A., {Georgantopoulos}, I., {et~al.} 2024, \mnras,
  527, 3674

\bibitem[{{Scargle} {et~al.}(2013){Scargle}, {Norris}, {Jackson}, \&
  {Chiang}}]{2013ApJ...764..167S}
{Scargle}, J.~D., {Norris}, J.~P., {Jackson}, B., \& {Chiang}, J. 2013, \apj,
  764, 167

\bibitem[{{Shapley}(1953)}]{shapley1953value}
{Shapley}, L.~S. 1953, in {The Shapley Value} (Princeton University Press,
  Princeton)

\bibitem[{{Shwartz-Ziv} \& {Armon}(2021)}]{2021arXiv210603253S}
{Shwartz-Ziv}, R. \& {Armon}, A. 2021, arXiv e-prints, arXiv:2106.03253

\bibitem[{{Sidoli} {et~al.}(2019){Sidoli}, {Postnov}, {Belfiore}, {Marelli},
  {Salvetti}, {Salvaterra}, {De Luca}, \& {Esposito}}]{2019MNRAS.487..420S}
{Sidoli}, L., {Postnov}, K.~A., {Belfiore}, A., {et~al.} 2019, \mnras, 487, 420

\bibitem[{{Song} {et~al.}(2025){Song}, {Villar}, {Martinez-Galarza}, \&
  {Dillmann}}]{2025arXiv250201627S}
{Song}, Y., {Villar}, V.~A., {Martinez-Galarza}, J.~R., \& {Dillmann}, S. 2025,
  arXiv e-prints, arXiv:2502.01627

\bibitem[{{Str{\"u}der} {et~al.}(2001){Str{\"u}der}, {Briel}, {Dennerl},
  {Hartmann}, {Kendziorra}, {Meidinger}, {Pfeffermann}, {Reppin}, {Aschenbach},
  {Bornemann}, {Br{\"a}uninger}, {Burkert}, {Elender}, {Freyberg}, {Haberl},
  {Hartner}, {Heuschmann}, {Hippmann}, {Kastelic}, {Kemmer}, {Kettenring},
  {Kink}, {Krause}, {M{\"u}ller}, {Oppitz}, {Pietsch}, {Popp}, {Predehl},
  {Read}, {Stephan}, {St{\"o}tter}, {Tr{\"u}mper}, {Holl}, {Kemmer}, {Soltau},
  {St{\"o}tter}, {Weber}, {Weichert}, {von Zanthier}, {Carathanassis}, {Lutz},
  {Richter}, {Solc}, {B{\"o}ttcher}, {Kuster}, {Staubert}, {Abbey}, {Holland},
  {Turner}, {Balasini}, {Bignami}, {La Palombara}, {Villa}, {Buttler},
  {Gianini}, {Lain{\'e}}, {Lumb}, \& {Dhez}}]{2001A&A...365L..18S}
{Str{\"u}der}, L., {Briel}, U., {Dennerl}, K., {et~al.} 2001, \aap, 365, L18

\bibitem[{{Turner} {et~al.}(2001){Turner}, {Abbey}, {Arnaud}, {Balasini},
  {Barbera}, {Belsole}, {Bennie}, {Bernard}, {Bignami}, {Boer}, {Briel},
  {Butler}, {Cara}, {Chabaud}, {Cole}, {Collura}, {Conte}, {Cros}, {Denby},
  {Dhez}, {Di Coco}, {Dowson}, {Ferrando}, {Ghizzardi}, {Gianotti}, {Goodall},
  {Gretton}, {Griffiths}, {Hainaut}, {Hochedez}, {Holland}, {Jourdain},
  {Kendziorra}, {Lagostina}, {Laine}, {La Palombara}, {Lortholary}, {Lumb},
  {Marty}, {Molendi}, {Pigot}, {Poindron}, {Pounds}, {Reeves}, {Reppin},
  {Rothenflug}, {Salvetat}, {Sauvageot}, {Schmitt}, {Sembay}, {Short},
  {Spragg}, {Stephen}, {Str{\"u}der}, {Tiengo}, {Trifoglio}, {Tr{\"u}mper},
  {Vercellone}, {Vigroux}, {Villa}, {Ward}, {Whitehead}, \&
  {Zonca}}]{2001A&A...365L..27T}
{Turner}, M.~J.~L., {Abbey}, A., {Arnaud}, M., {et~al.} 2001, \aap, 365, L27

\bibitem[{{Yang} {et~al.}(2019){Yang}, {Brandt}, {Zhu}, {Bauer}, {Luo}, {Xue},
  \& {Zheng}}]{2019MNRAS.487.4721Y}
{Yang}, G., {Brandt}, W.~N., {Zhu}, S.~F., {et~al.} 2019, \mnras, 487, 4721

\bibitem[{{Yang} {et~al.}(2024){Yang}, {Hare}, \&
  {Kargaltsev}}]{2024ApJ...971..180Y}
{Yang}, H., {Hare}, J., \& {Kargaltsev}, O. 2024, \apj, 971, 180

\bibitem[{{Zhang} {et~al.}(2021){Zhang}, {Zhao}, \& {Wu}}]{2021MNRAS.503.5263Z}
{Zhang}, Y., {Zhao}, Y., \& {Wu}, X.-B. 2021, \mnras, 503, 5263

\bibitem[{{Zuo} {et~al.}(2024){Zuo}, {Tao}, {Liu}, {Xu}, {Zhang}, {Pan}, {Sun},
  {Zhang}, {Cui}, \& {Yuan}}]{2024RAA....24h5016Z}
{Zuo}, X., {Tao}, Y., {Liu}, Y., {et~al.} 2024, Research in Astronomy and
  Astrophysics, 24, 085016

\end{thebibliography}

\begin{appendix}
\section{Data analysis and labels definition}
\label{label}

As described in Section \ref{intro}, recently we extended the search for aperiodic variability of {\it XMM-Newton/EPIC} sources that is part of the EXTraS project \citep{2021A&A...650A.167D}. While a complete description of the work will be presented in a future paper (Marelli et al. in prep.) and the results will be publicly available through a dedicated site (\url{https://www.iasf-milano.inaf.it/extras/}), we report here the main implementations that concern this work.

The new EXTraS-Aperiodic project extends the original data set of EXTraS to all the public {\it XMM-Newton} observation from its launch to the end of 2020.\\
Apart from minor changes, e.g. script optimization and bug fixes, the exposure light curves (LC) extraction is basically the same, taking as inputs the raw ODF {\it XMM-Newton} data and the sources and their parameters from the XMM serendipitous source catalog\footnote{http://xmm-catalog.irap.omp.eu/}.

We merge the (uniform bin) LC in count rate from different cameras (pn and the two MOSs) into a single LC in flux. For each time bin, the count rate was converted to 0.2–12 keV flux using information from the XMM-Newton source catalog. The contemporaneous bins from the three cameras were then merged by computing a weighted mean, with the associated uncertainties propagated accordingly. This allows us to take into account the different characteristics of each camera to produce a consistent LC for the entire observation, even in the periods in which one or more cameras are off.

\begin{figure}
\includegraphics[width=\columnwidth]{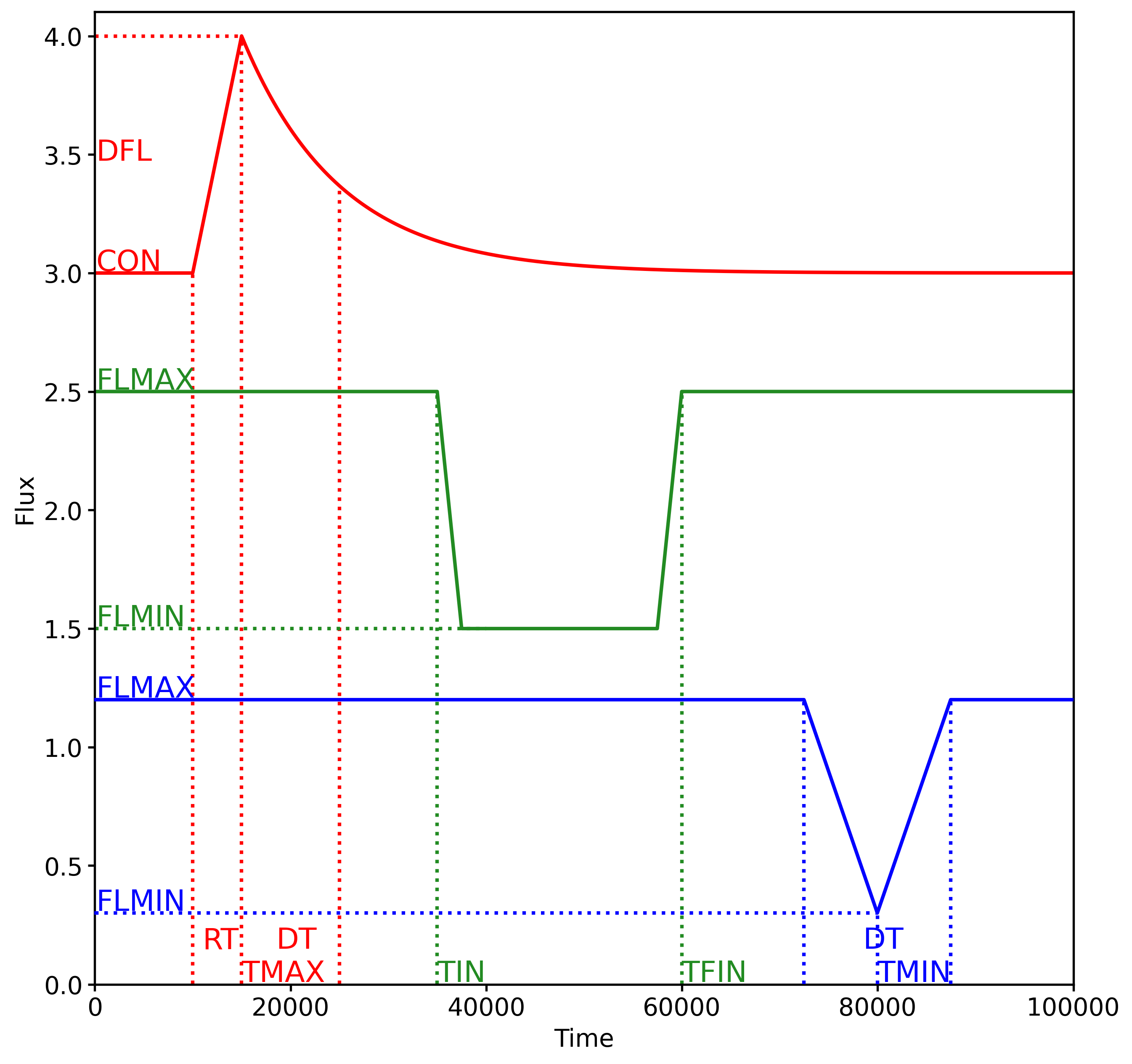}
\caption{This image shows an example of three of the models we fitted, with the associated features. We show in red the flare model, in green the single eclipse model and in blue the single dip model. \label{figmodels}}
\end{figure}

From the merged LC we extract 130 parameters and flags 
to characterize their variability. These can be divided into two main groups: the model-independent statistical features and the model-dependent features. The former include standard statistical features like the weighted average, standard deviation, skewness, and kurtosis; most of them are listed in table \ref{tab-features}. The latter include the goodness and best fit parameters for a number of models; in this case also most of them are listed in table \ref{tab-features}. We fitted constant, linear, quadratic and exponential models. We also implemented some more complex models in order to search in the LC for features of astrophysical interest, such as flares, dips (single and periodic) and eclipses (single and periodic).
Simple visualizations of the features of these models are shown in Figure \ref{figmodels}.
In addition to the features from the models themselves, we also compared the results of different models using f-tests \citep{1969drea.book.....B}, in particular comparing more complex models to a constant fit. We run a comparison of the likelihood for nested models, using the f-test. The asymptotic distribution under the null hypothesis that the nested model is correct was then converted in sigma units. This way, the statistical gain across all sources has a compact distribution, more robust towards outliers.

\begin{figure}
\begin{tabular}{c}
\includegraphics[width=0.96\columnwidth]{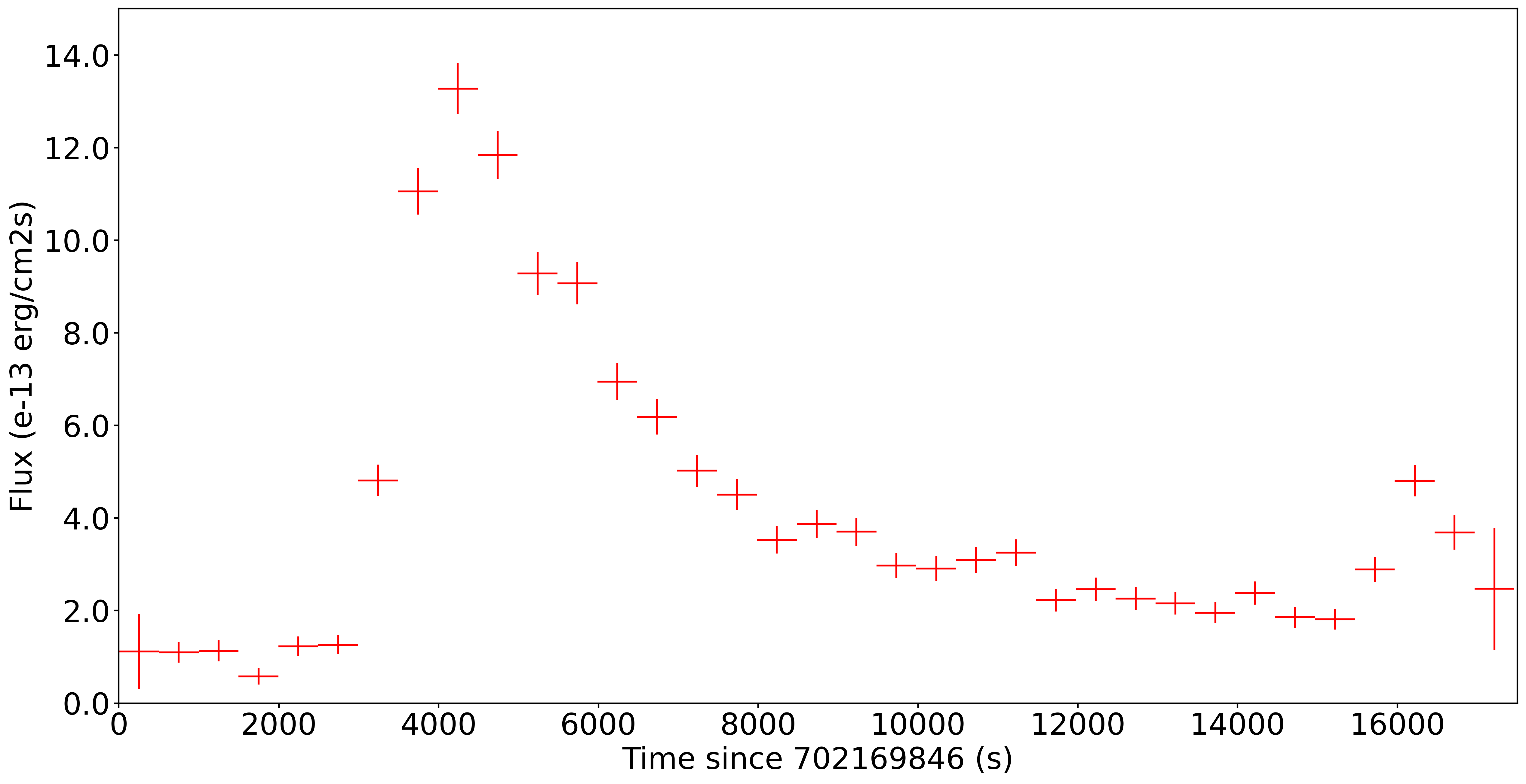}\\
\includegraphics[width=\columnwidth]{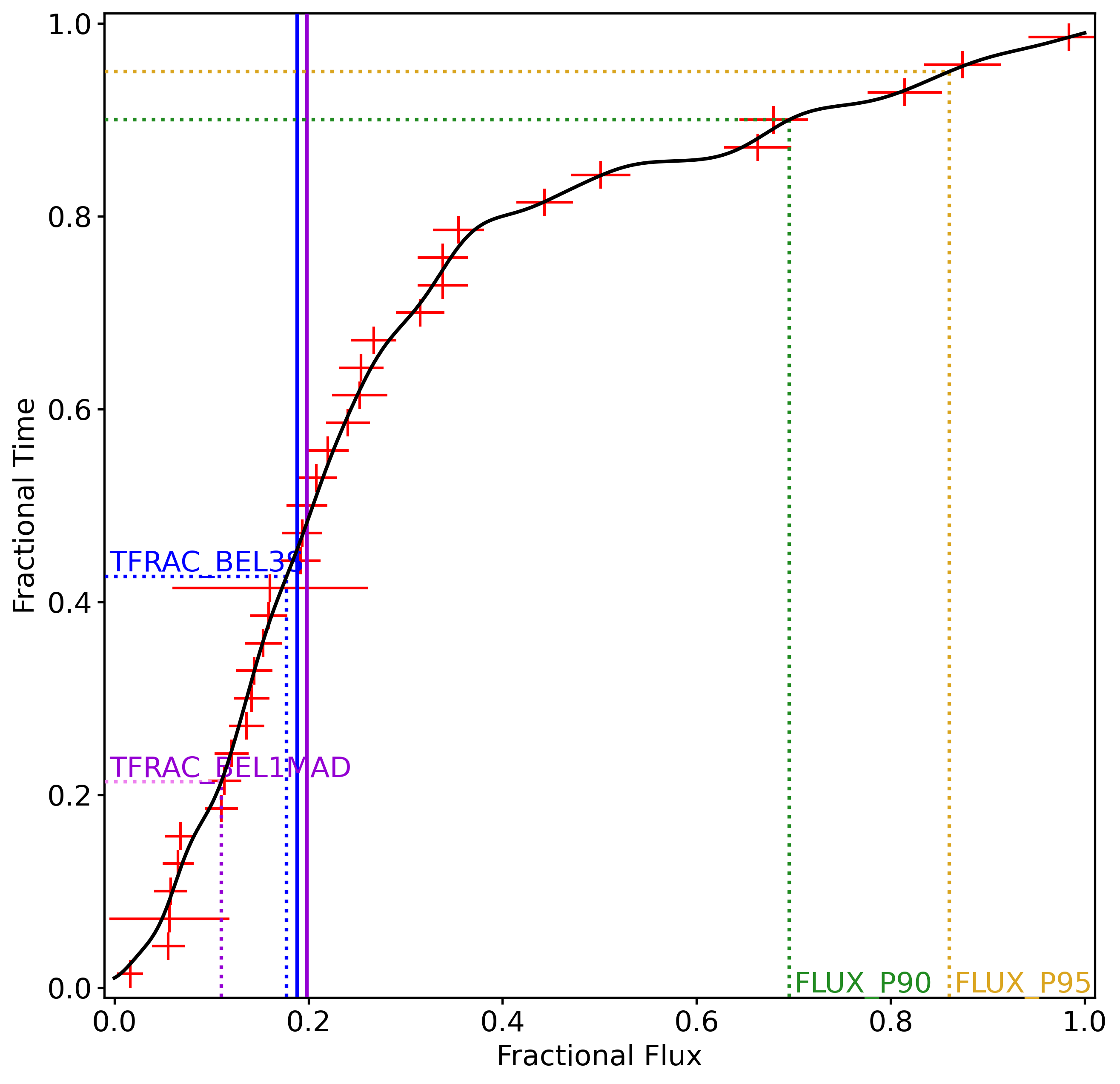}\\
\end{tabular}
\caption{Example of a flaring LC ({\it Upper Panel}) with the associated CDF ({\it Lower Panel}). In red we show the original discrete CDF, in black we show the continuous CDF; in blue and violet we show respectively the weighted average and the median of the LC. We also show some of the CDF features we use: TFRAC\_BEL1MAD, TFRAC\_BEL3S, FLUX\_90 and FLUX\_95 (respectively, in violet, blue, green and orange).\label{figcdf}}
\end{figure}

Based on the results of many papers in the literature \citep[see e.g.][]{2011ApJ...733...10R,2014ApJ...786...20L,2015ApJ...813...28F}, the cumulative distribution function (CDF) offers several features that are useful for describing the associated light curve (LC) variability. Thus, from each merged LC, we extracted the corresponding CDF.\\
Unlike what we did in the original EXTraS, we extracted a continuous, smoothed CDF. Each bin is treated as a normalized (by its integral) Gaussian on the flux axis of the CDF, with the Gaussian $\mu$ at the bin value and the (2x)bin error as its $\sigma$; the Gaussians are then merged to produce a continuous CDF. Once smoothed, we normalize the flux axis so that the flux at 1\% of the cumulative time is 0 and the flux at 99\% of the cumulative time is 1; this way, the shape of the CDF (and thus the extracted features) is both time and flux independent. We show an example in Figure \ref{figcdf}.

From each smoothed CDF, we extracted 53 features to describe its shape (plus 3 for the normalization), each one being the normalized flux at (or flux range between) certain cumulative time(s) or the cumulative time at certain normalized flux. Our features are chosen in order to reproduce the ones reported in \cite{2011ApJ...733...10R,2014ApJ...786...20L,2015ApJ...813...28F} through a simple linear combination. Most of these features are reported in Table \ref{tab-features}.

\newpage
{\small
\tablefirsthead{\hline\hline - & Name & Description \\ \hline}
\begin{supertabular}{lll}
1 & N\_BINS & Number of time bins\\
2 & STDEV & Standard deviation\\
3 & SKEW & Skewness\\
4 & KURT & Kurtosis\\
5 & AMPLIT & Amplitude\\
  &        & $(max(LC)-min(LC)/2$\\
6 & MEDABSDEV & MAD, Median deviation from median\\
  &        & $median(|LC-median(LC)|)$\\
7 & MEDMAXOFF & Max deviation from median on median\\
  &        & $max(|LC-median(LC)|)/|median(LC)|$ \\
\hline
8 & CON\_NSIGMA & Inverse survival function (constant model)\\ 
9 & LIN\_NSIGMA & Inverse survival function (linear model)\\
10 & LIN\_CON & Constant term (linear model)\\
11 & LIN\_CON\_E$^{a}$ & $1\sigma$ error in LIN\_CON\\ 
12 & LIN\_LIN & Linear term (linear model)\\
13 & LIN\_LIN\_E$^{a}$ & $1\sigma$ error on LIN\_LIN\\
14 & QU\_NSIGMA & Inverse survival function (quadratic model)\\
15 & QU\_CON & Constant term (quadratic model)\\
16 & QU\_CON\_E$^{a}$ & $1\sigma$ error on QU\_CON\\
17 & QU\_LIN & Linear term (quadratic model)\\
18 & QU\_LIN\_E$^{a}$ & $1\sigma$ error on QU\_LIN\\
19 & QU\_QU & Quadratic term (quadratic model)\\
20 & QU\_QU\_E$^{a}$ & $1\sigma$ error on QU\_QU\\
21 & DIP1\_NSIGMA & Inverse survival function (s.dip model)\\
22 & DIP1\_TMIN & Time of the minimum (s.dip model)\\
23 & DIP1\_TMIN\_E$^{a}$ & $1\sigma$ error on DIP1\_TMIN\\
24 & DIP1\_DT & Time duration of the dip (s.dip model)\\
25 & DIP1\_DT\_E$^{a}$ & $1\sigma$ error on DIP1\_DT\\
26 & DIP1\_FLMIN & Flux ad the minimum (s.dip model)\\
27 & DIP1\_FLMIN\_E$^{a}$ & $1\sigma$ error on DIP1\_FLMIN\\
28 & DIP1\_FLMAX & Constant term (s.dip model)\\
29 & DIP1\_FLMAX\_E$^{a}$ & $1\sigma$ error on DIP1\_FLMAX\\
30 & DIP\_NSIGMA & Inverse survival function (m.dip model)\\
31 & DIP\_TMIN & Phase of the minimum (m.dip model)\\
32 & DIP\_TMIN\_E$^{a}$ & $1\sigma$ error on DIP\_TMIN\\
33 & DIP\_DT & Phase duration of the dip (m.dip model)\\
34 & DIP\_DT\_E$^{a}$ & $1\sigma$ error on DIP\_DT\\
35 & DIP\_FLMIN & Flux ad the minimum (m.dip model)\\
36 & DIP\_FLMIN\_E$^{a}$ & $1\sigma$ error on DIP\_FLMIN\\
37 & DIP\_FLMAX & Constant term (m.dip model)\\
38 & DIP\_FLMAX\_E$^{a}$ & $1\sigma$ error on DIP\_FLMAX\\
39 & DIP\_PER & Time period (m.dip model)\\
40 & DIP\_PER\_E$^{a}$ & $1\sigma$ error on DIP\_PER\\
41 & EC1\_NSIGMA & Inverse survival function (s.ecl. model)\\
42 & EC1\_TIN & Begin time of eclipse (s.ecl. model)\\
43 & EC1\_TIN\_E$^{a}$ & $1\sigma$ error on EC1\_TIN\\
44 & EC1\_TFIN & End time of eclipse (s.ecl. model)\\
45 & EC1\_TFIN\_E$^{a}$ & $1\sigma$ error on EC1\_TFIN\\
46 & EC1\_FLMIN & Flux in the eclipse (s.ecl. model)\\
47 & EC1\_FLMIN\_E$^{a}$ & $1\sigma$ error on EC1\_FLMIN\\
48 & EC1\_FLMAX & Constant term (s.ecl. model)\\
49 & EC1\_FLMAX\_E$^{a}$ & $1\sigma$ error on EC1\_FLMAX\\
50 & EC\_NSIGMA & Inverse survival function (m.ecl. model)\\
51 & EC\_T1 & Begin phase of eclipse (m.ecl. model)\\
52 & EC\_T1\_E$^{a}$ & $1\sigma$ error on EC\_T1\\
53 & EC\_T2 & End phase of eclipse (m.ecl. model)\\
54 & EC\_T2\_E$^{a}$ & $1\sigma$ error on EC\_T2\\
55 & EC\_FL1 & Flux level 1 (m.ecl. model)\\ 
56 & EC\_FL1\_E$^{a}$ & $1\sigma$ error on EC\_FL1\\
57 & EC\_FL2 & Flux level 2 (m.ecl.model)\\
58 & EC\_FL2\_E$^{a}$ & $1\sigma$ error on EC\_FL2\\
59 & EC\_PER & Time period (m.ecl. model)\\
60 & EC\_PER\_E$^{a}$ & $1\sigma$ error on EC\_PER\\
61 & FL\_NSIGMA & Inverse survival function (flare model)\\
62 & FL\_CON & Constant term (flare model)\\
63 & FL\_CON\_E$^{a}$ & $1\sigma$ error on FL\_CON\\
64 & FL\_DFL & $\Delta$Flux at maximum (flare model)\\
65 & FL\_DFL\_E$^{a}$ & $1\sigma$ error on FL\_DFL\\
66 & FL\_TMAX & Time of maximum (flare model)\\
67 & FL\_TMAX\_E$^{a}$ & $1\sigma$ error on FL\_TMAX\\
68 & FL\_DT & Decay time (flare model)\\
69 & FL\_DT\_E$^{a}$ & $1\sigma$ error on FL\_DT\\
70 & FL\_RT & Rising time (flare model)\\
71 & FL\_RT\_E$^{a}$ & $1\sigma$ error on FL\_RT\\
72 & F\_NSIGMA\_LINCON & f-test prob. in $\sigma$s (linear-const.)\\ 
73 & F\_NSIGMA\_QUCON & f-test prob. in $\sigma$s (quadratic-const.)\\
74 & F\_NSIGMA\_QULIN & f-test prob. in $\sigma$s (quadratic-linear)\\
75 & F\_NSIGMA\_DIP1CON & f-test prob. in $\sigma$s (s.dip-const.)\\
76 & F\_NSIGMA\_DIPCON & f-test prob. in $\sigma$s (m.dip-const.)\\
77 & F\_NSIGMA\_DIP1DIP & f-test prob. in $\sigma$s (m.dip-s.dip)\\
78 & F\_NSIGMA\_EC1CON & f-test prob. in $\sigma$s (s.ecl.-const.)\\
79 & F\_NSIGMA\_ECCON & f-test prob. in $\sigma$s (m.ecl.-const.)\\
80 & F\_NSIGMA\_EC1EC & f-test prob. in $\sigma$s (m.ecl.-s.ecl.)\\
81 & F\_NSIGMA\_FLCON & f-test prob. in $\sigma$s (flare-const.)\\
\hline
82 & TFRAC\_MID20 & \%time between 0.9-1.1 of MEDIAN\\
83 & TFRAC\_BEL1S & \%time below AVE-AVE\_ERR$^{b}$\\
84 & TFRAC\_ABO1S & \%time above AVE+AVE\_ERR$^{b}$\\
85 & TFRAC\_BEL3S & \%time below AVE-3AVE\_ERR$^{b}$\\
86 & TFRAC\_ABO3S & \%time above AVE+3AVE\_ERR$^{b}$\\
87 & TFRAC\_BEL5S & \%time below AVE-5AVE\_ERR$^{b}$\\
88 & TFRAC\_ABO5S & \%time frac. above AVE+5AVE\_ERR$^{b}$\\
89 & TFRAC\_BEL1MAD & \%time below MEDIAN-MAD$^{c}$\\
90 & TFRAC\_ABO1MAD & \%time above MEDIAN+MAD$^{c}$\\
91 & TFRAC\_BEL3MAD & \%time below MEDIAN-3MAD$^{c}$\\
92 & TFRAC\_ABO3MAD & \%time above MEDIAN+3MAD$^{c}$\\
93 & TFRAC\_BEL5MAD & \%time below MEDIAN-5MAD$^{c}$\\
94 & TFRAC\_ABO5MAD & \%time above MEDIAN+5MAD$^{c}$\\
95 & FLUX\_P05 & \%flux at 5\% of cumulative time\\ 
96 & FLUX\_P10 & \%flux below 10\% of cumulative time\\
97 & FLUX\_P17.5 & \%flux below 17.5\% of cumulative time\\
98 & FLUX\_P25 & \%flux below 25\% of cumulative time\\
99 & FLUX\_P32.5 & \%flux below 32.5\% of cumulative time\\
100 & FLUX\_P40 & \%flux below 40\% of cumulative time\\
101 & FLUX\_P50 & \%flux below 50\% of cumulative time\\
102 & FLUX\_P60 & \%flux below 60\% of cumulative time\\
103 & FLUX\_P67.5 & \%flux below 76.5\% of cumulative time\\
104 & FLUX\_P75 & \%flux below 75\% of cumulative time\\
105 & FLUX\_P82.5 & \%flux below 82.5\% of cumulative time\\
106 & FLUX\_P90 & \%flux below 90\% of cumulative time\\
107 & FLUX\_P95 & \%flux below 95\% of cumulative time\\
108 & MEDIAN\_01 & Normalized MEDIAN\\ 
\hline
\hline
\end{supertabular}
\captionof{table}{We show the complete list of the features we used, with a brief explanation for each one and its formula, if necessary. We divided the list in three sublists separated by horizontal lines to show the model-independent statistical features, model-dependent features and CDF features, respectively.\\
$^{a}$ \_ERR has been written as \_E to fit the table.\\
$^{b}$ Weighted average and its $1\sigma$ error.\\
$^{c}$ MEDIAN and MEDABSDEV.}
\label{tab-features}
}

\section{UMAP plots for the most important features}
\label{appendicebonus}
\begin{figure*}
\begin{tabular}{cc}
\includegraphics[width=\columnwidth]{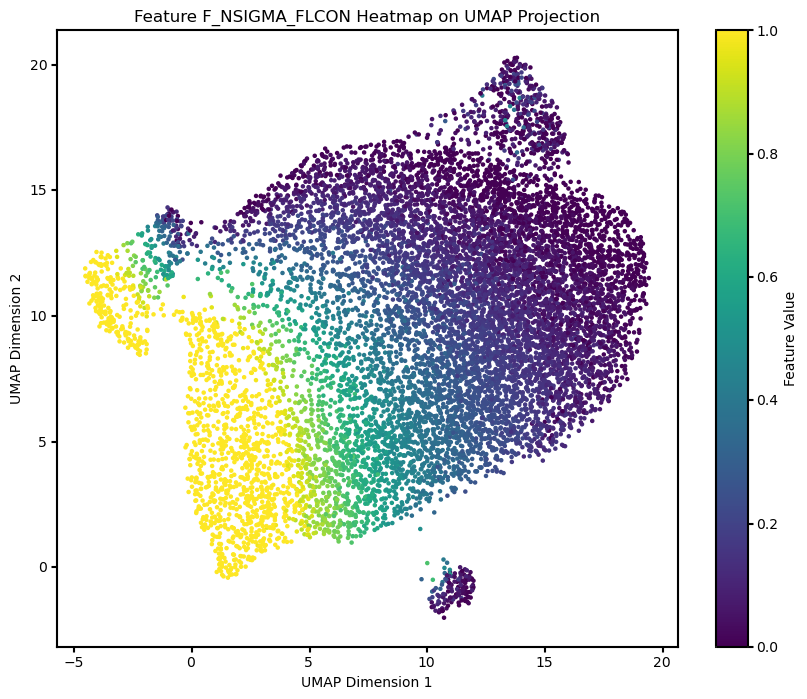} & \includegraphics[width=\columnwidth]{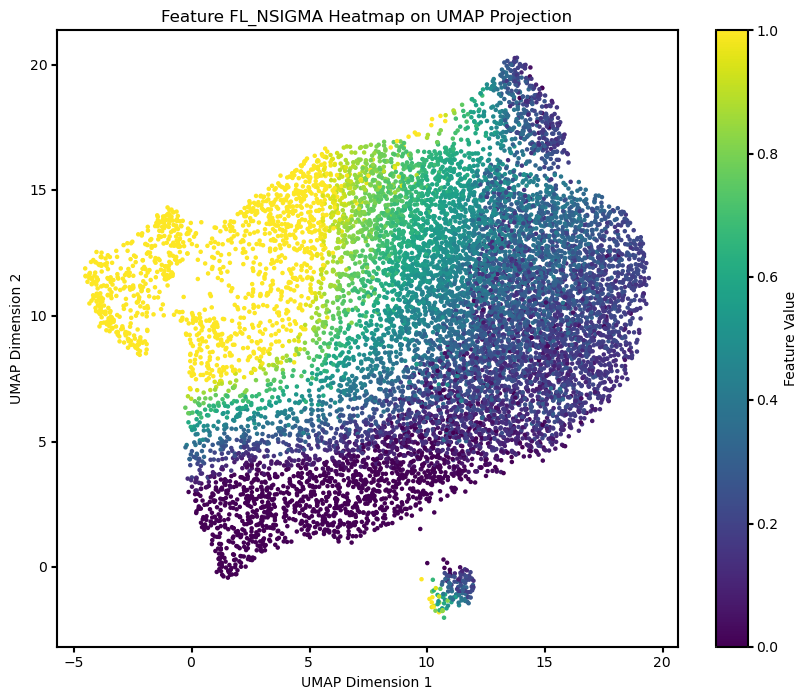}\\
\includegraphics[width=\columnwidth]{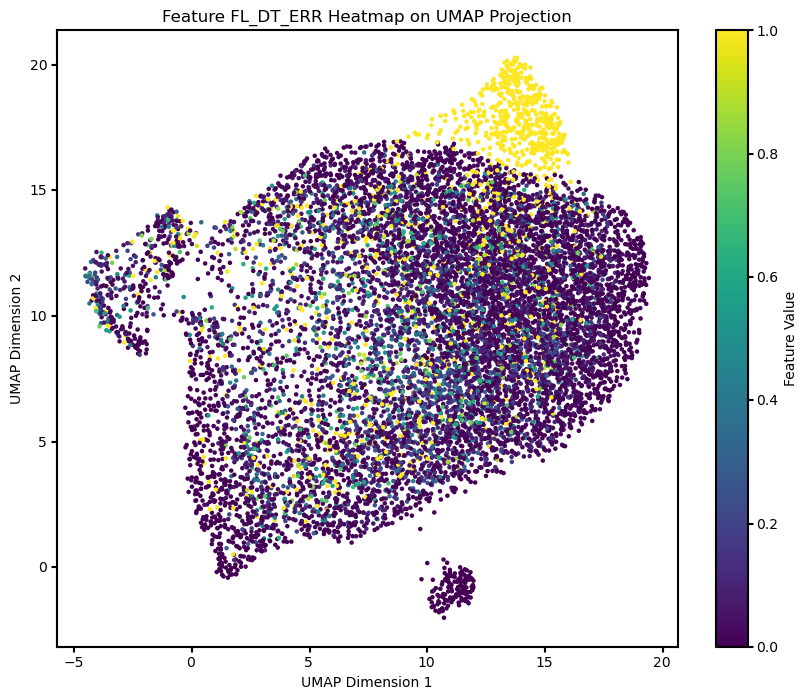} & \includegraphics[width=\columnwidth]{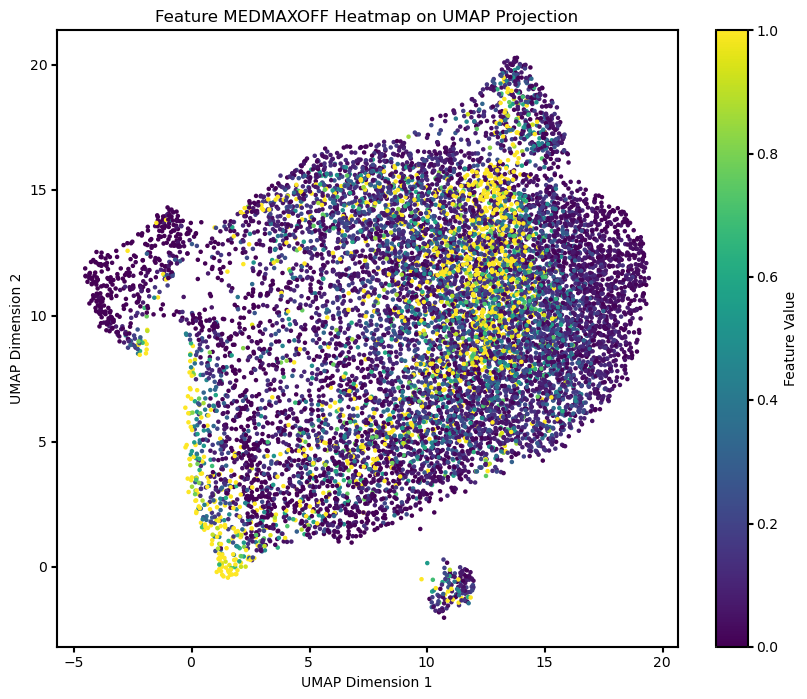}\\
\includegraphics[width=\columnwidth]{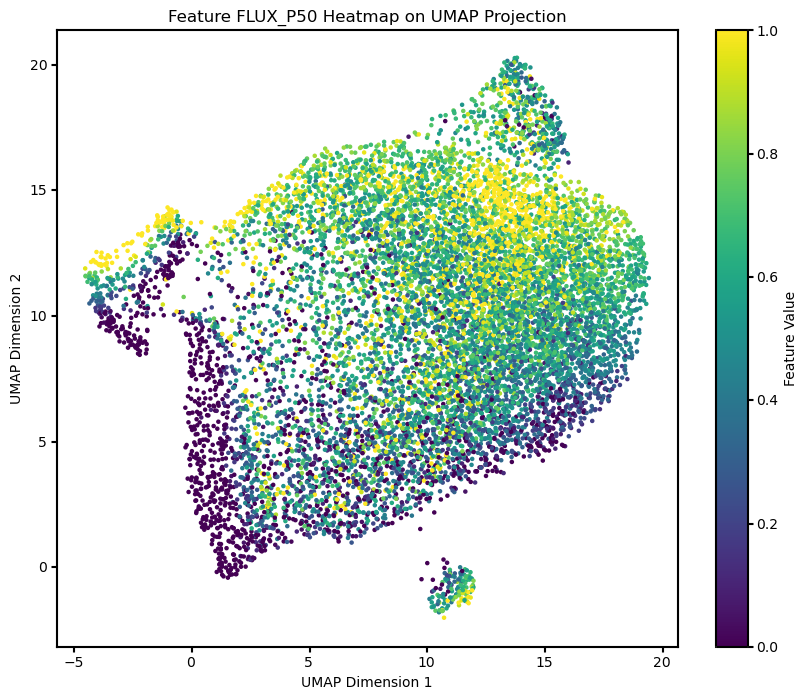} & \includegraphics[width=\columnwidth]{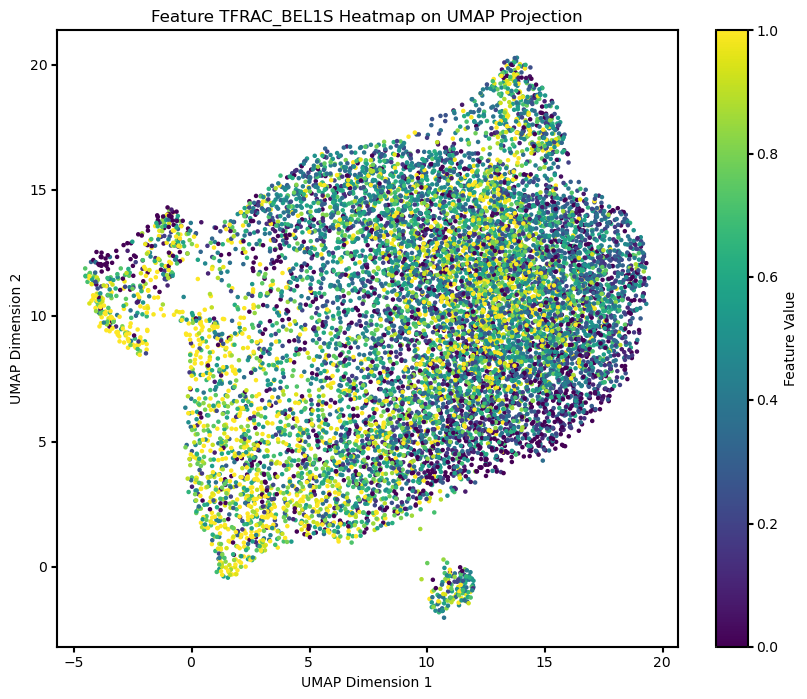}\\
\end{tabular}
\caption{Feature by feature visualization on the UMAP plane.\label{umapappe}}
\end{figure*}

Here we report (Fig.~\ref{umapappe}) the detailed UMAP views for each one of the six most important features. For interpretation see Sect.~\ref{appe}.

\label{lastpage}
\end{appendix}
\end{document}